\documentclass{article}
\usepackage{amsmath,amsfonts,graphicx} 
\usepackage{amssymb,mathtools}
\usepackage[letterpaper, margin=1in]{geometry}

\usepackage{soul}

\usepackage[T1]{fontenc}
\usepackage{newpxtext,newpxmath}

\usepackage{authblk}
\usepackage{hyperref}
\usepackage{orcidlink}

\newcommand{\orcid}[1]{\orcidlink{#1}}
\usepackage{xcolor}

 \usepackage[numbers,sort&compress]{natbib}

\usepackage{caption}
\numberwithin{equation}{section}

\def\pr#1{\text{Pr}\left(#1\right)}

\usepackage{soul}

\title{Daring few, patient many: division of labor in decentralized foraging collectives}

\author[1,*]{\orcid{0000-0002-3534-2102}Hyunjoong Kim}
\author[2,3,*]{\orcid{0000-0002-2835-9416}Zachary P Kilpatrick}
\author[4,5,6,*]{\orcid{0000-0002-1975-3913}Kre\v{s}imir Josi\'c}
\affil[1]{{\small Department of Mathematical Sciences, University of Cincinnati, Cincinnati, OH 45221}}
\affil[2]{{\small Department of Applied Mathematics, University of Colorado Boulder, Boulder, CO 80309}}
\affil[3]{{\small Institute of Cognitive Science, University of Colorado Boulder, Boulder, CO 80309}}
\affil[4]{{\small Department of Mathematics, University of Houston, Houston, TX 77204}}
\affil[5]{{\small Department of Biology and Biochemistry, University of Houston, Houston, TX 77204}}
\affil[6]{{\small NSF-Simons National Institute for Theory and Mathematics in Biology, Chicago, IL 60611}}
\affil[*]{{\small \href{mailto:hyunjoong.kim@uc.edu}{\texttt{hyunjoong.kim@uc.edu}}, \href{mailto:zpkilpat@colorado.edu}{\texttt{zpkilpat@colorado.edu}}, \href{mailto:kresimir.josic@gmail.com}{\texttt{kresimir.josic@gmail.com}}}}

\begin{document}

\maketitle

\begin{abstract}
How do social animals make effective decisions in the absence of a leader? While coordination can improve accuracy, it also introduces delays as information propagates through the group. In changing environments, these delays can outweigh the benefits of globally coordinated decisions, even when local interactions remain tightly organized. This raises a key question: how can groups implement efficient collective decision-making without central coordination?
We address this question using a collective foraging model in which individuals share information and rewards, but each must choose whether to bear the cost of exploring or to remain idle. We show that decentralized collectives can match the performance of centrally controlled groups through a division of labor: a small, heterogeneous subset explores even when expected rewards are negative, acquiring information to enable future foraging, while a coordinated majority forages only when expected rewards are positive. Information redundancy causes the optimal number of explorers to grow sublinearly with group size, so that larger groups need proportionally fewer explorers. The heterogeneity of the group is maximized at intermediate ecological pressures, but optimal groups are homogeneous when costs or fluctuations are extreme. 
Crucially, these group-level policies do not require central coordination, emerging instead from agents following simple threshold-based decision rules. We thus demonstrate a mechanism through which leaderless collectives can make effective decisions under uncertainty and show how ecological pressures can drive changes in the distribution of strategies employed by the group. 
\end{abstract}


\newpage
\section{Introduction}
How animal groups balance individual exploration costs and collective information gains remains a central question in understanding ecological decision-making \cite{Conradt2003,Couzin2005,Sumpter2005,Couzin2009}.
Individual foragers bear energetic and survival costs when venturing into uncertain environments \cite{Stephens1986,Wolf1989}, yet the information about the environmental state resulting from these forays can outweigh any immediate rewards \cite{Couzin2005,Hein2015,Dall2005}.
This tension raises a fundamental question: 
How should groups be organized to balance these costs and benefits?

One solution to this problem is centralized leadership, in which a small number of individuals guide collective action. 
In many wild wolf packs, for example, the breeding pair tends to initiate movements and coordinates traveling, resting, and hunting~\cite{Mech1999}. 
Human history offers parallel examples. 
Admiral Zheng He commanded fleets across the Indian Ocean in the early fifteenth century on imperial orders, spurring Chinese emigration and trade throughout Southeast Asia~\cite{Levathes2014}. A century later, royal sponsorship supported Magellan's circumnavigation, opening the first western sea route to the Spice Islands and accelerated global trade and contact between distant regions~\cite{Bergreen2003}.
In each case, a small number of designated explorers, acting under a central authority, bore enormous personal risk, while the societies that sent them reaped lasting rewards.
However, most social animals, including honeybees, ants, and fish schools, lack fixed leaders and coordinate through local interactions~\cite{Pratt2002,Seeley2004,Couzin2005,Sumpter2005,Seeley2011,Hein2015}.
Despite extensive empirical study of decentralized animal groups~\cite{Pratt2002,Sumpter2005,Giardina2008,Gordon2014}, 
it remains unclear how closely decentralized coordination can match centralized control, how performance depends on diversity in individual strategies, and under what conditions heterogeneous strategies offer an advantage in leaderless groups.

To address these questions, we develop and analyze a stochastic foraging model~\cite{Puterman1994,Mahadevan1996,Gold2007,Kilpatrick2021,BlumMoyse2025,Dembo2010,Bressloff2021,Barendregt2025} to understand how collectives can achieve efficient exploration. We show that well-calibrated collectives exhibit two characteristic behavioral regimes: during times of plenty, risk-averse individuals venture out together, amplifying collective gains. 
When resources are scarce, most individuals refrain from foraging, but a small subset of risk-tolerant explorers continues foraging to rapidly detect environmental improvements. Composition is critical as populations dominated by cautious individuals fail to discover new resources, whereas those dominated by risk-tolerant foragers exhaust shared reserves. The number of risk-tolerant individuals grows sublinearly with population size, making a balanced division of labor between exploration and exploitation essential for efficient foraging~\cite{Seeley2011,Charbonneau2015}.
Remarkably, collectives with a well-calibrated mixture of cautious and bold individuals can perform as well as groups governed by a central decision-maker. 
The most heterogeneous compositions emerge under intermediate ecological pressures, when exploration costs and environmental variability are balanced.  We thus show how simple individual decision rules lead to optimal collective performance and predict when ecological pressures favor heterogeneous versus uniform group composition.

\begin{figure*}[b!]
    \centering
    \includegraphics[width=.99\linewidth]{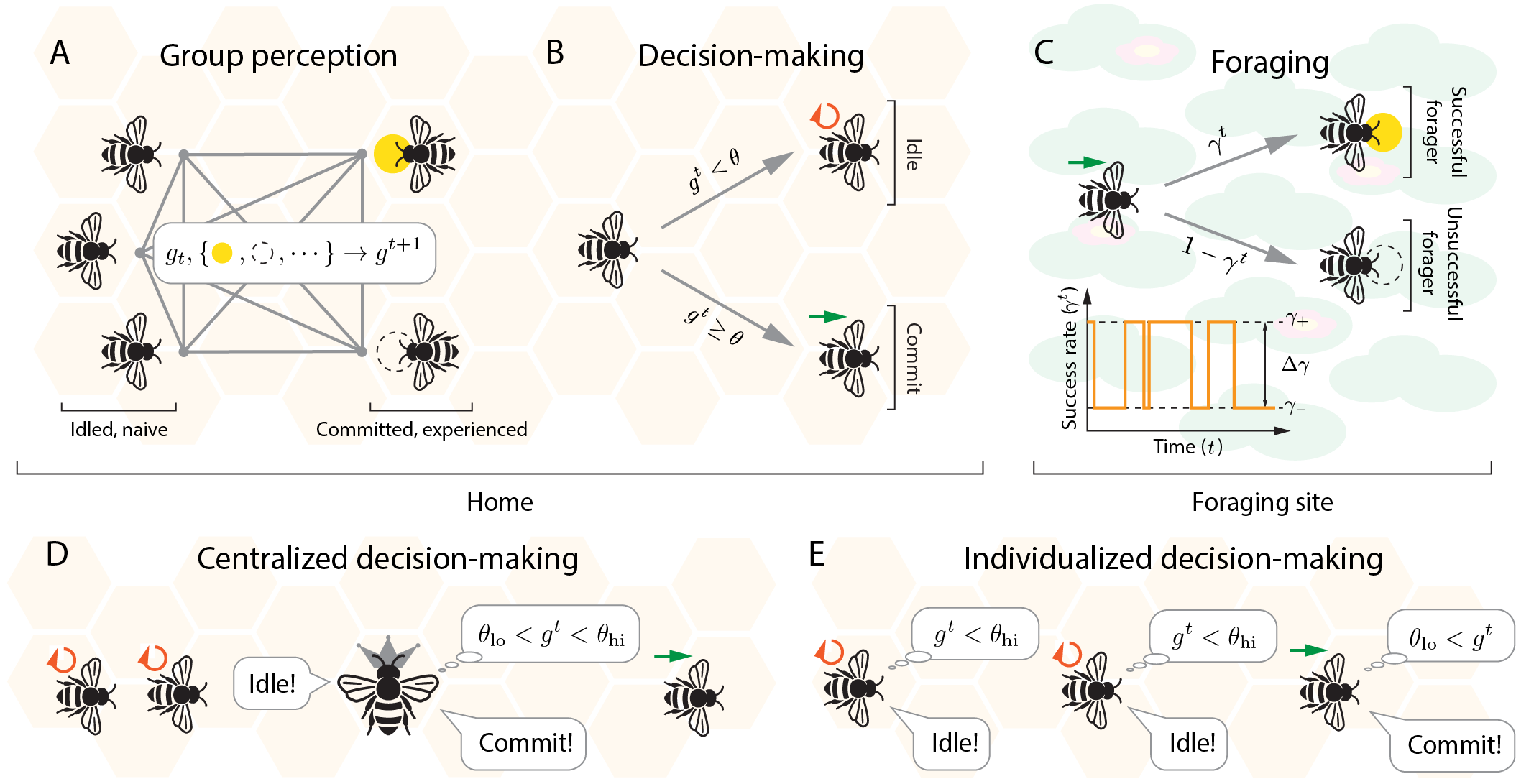}
    \caption{
    \textbf{Overview of model and setup.}
    Individuals in a collective observe returning foragers to update their belief about the environmental state and select actions. 
    \textbf{(A)}~The belief, $g^t$, is updated based on the observed outcomes of foraging attempts. 
    \textbf{(B)}~Individuals forage when their belief that the environment is in a high-reward state exceeds a threshold, $
    \theta$; these thresholds can differ between individuals. 
    \textbf{(C)}~Foraging is successful with probability $\gamma^t$, which switches between high and low values with the environmental state. 
    \textbf{(D)}~Under centralized control, a planner decides on the optimal number of foragers to maximize returns.
    \textbf{(E)}~With decentralized control, coordination emerges from individual foraging decisions.
    }
    \label{fig:model}
\end{figure*}

\section{A Model of Collective Foraging under Uncertainty}
Collective decisions about foraging in large, decentralized collectives face a fundamental challenge: how do groups balance individual risk and collective benefit in changing environments without central coordination?
To address this question, we consider a minimal model designed to examine whether decentralized decision-making mechanisms can yield optimal collective long-term returns, even when individual expected returns are negative.
Here we provide an overview of the model;  mathematical details  can be found in SI Appendix, Section 1.

\subsection*{Foraging agents in a dynamic environment}
We consider a collective of $N$ agents who forage for resources. Time is discrete, and on each time step an agent chooses either to leave the ``hive'' and forage, or to stay. Each foraging attempt is either successful, resulting in a positive reward, or unsuccessful, resulting in no reward. The environment alternates between a high- and low-rewarding state, so that in the high (low) state a foraging attempt is successful with probability $\gamma_+$ ($\gamma_-$).  To model energetic expenditure and environmental risk, we assume that each foraging attempt incurs a fixed cost, $\lambda$, and that the expected return from an attempt is positive in the high-reward state and negative in the low-reward state. 
Transitions between the states are memoryless, so the environmental state evolves as a two-state discrete-time Markov process with symmetric switching probability, $\epsilon$ (Fig.~\ref{fig:model}C). While real environments fluctuate across a continuum of states, assuming two states provides a minimal framework that captures the essential trade-off between exploration and exploitation under uncertainty~\cite{Gold2007,Kilpatrick2021,Barendregt2025,BlumMoyse2025}.

\subsection*{Collective perception}
We assume that the community coordinates its behavior through shared information about the outcomes of foraging attempts. 
Returning foragers are observed by the entire community, so all individuals know how many foraging attempts have been successful (Fig.~\ref{fig:model}A), an idealization of group-level information flow in social collectives~\cite{Sumpter2005,Sumpter2008,Couzin2005}.
Individuals are Bayesian observers \cite{Wald1948,Bogacz2006,VelizCuba2016} and compute the probability that the environmental state is in a high-reward state based on the observed number of successful foragers and knowledge about the environment's volatility. We refer to this subjective probability as their \emph{belief} about the environmental state. Because all individuals have access to the same information, they share a common belief that guides their foraging decisions.

\subsection*{Decision-making with a centralized planner}
Maximizing returns requires optimizing forager number: too many foragers when conditions are poor wastes effort; too few when conditions are good misses opportunities. To establish an upper bound on collective performance, we derive a theoretical benchmark based on a centralized planner that optimally determines foraging effort required to maximize long-term collective returns (Fig.~\ref{fig:model}D). Under this centralized policy, all individuals follow the directives of a planner, which infers the environmental state by observing foraging successes and then prescribes the best number of foragers. This optimal policy can be computed using dynamic programming \cite{Bellman1954,Mahadevan1996,Sutton1998} and provides an upper bound on achievable performance. We compare this benchmark with outcomes from agents who make individual decisions based on shared information but without central coordination.

\subsection*{Decisions in the absence of a centralized planner}
With no central decision-maker, each individual decides whether to forage based on their estimate of the environmental state. We assume that each agent chooses to forage if their belief that the environment is in the high-reward state exceeds a private threshold (Fig.~\ref{fig:model}B, E). The distribution of these thresholds across individuals determines the behavior of the decentralized group~\cite{Bonabeau1996,Bonabeau1998}.
Such threshold-based decision rules are consistent with empirical work on quorum/acceptability thresholds in social insects and with theory showing how heterogeneity in thresholds shapes collective decisions~\cite{Theraulaz1998,Franks2003,Robinson2011,Conradt2012,Jeanson2014,Masuda2015}.

\vspace{3mm}
\noindent Despite its abstraction, this model captures the ecological tradeoffs between exploration cost, environmental uncertainty, and collective return. It also allows us to ask 
how well a decentralized collective can perform compared to an optimally controlled group.

\section{Results}

Using our model of foraging and belief updating, we ask how a collective translates shared environmental information into effective foraging decisions.
We first establish a normative benchmark based on a centralized decision-maker that maximizes expected payoff.
We then show that a leaderless group can achieve optimal performance using heterogeneous individual decision thresholds and analyze how environmental conditions shape the optimal threshold distribution in a decentralized collective.

\subsection{Robust collective coordination through structured heterogeneity}

A central decision-maker can select an optimal number of foragers to maximize expected payoffs on each turn.
In a decentralized collective, individuals share a common belief about the environmental state but make individual choices based on a private threshold. If all individuals choose a threshold that maximizes their own expected return, then all thresholds are identical. Since they share the same belief, everyone makes the same decision: forage when the expected reward is positive, and remain idle when the expected reward is negative. As a result, a collective can quickly detect when an environment switches from a high- to a low-rewarding state. However, when the collective believes the reward state is likely to be low, no individuals forage, and the collective will not detect a switch back to a high-rewarding state. The resulting failure to explore thus leads to starvation even when conditions improve. 
Conversely, if all individuals adopt low thresholds and forage often, the collective can quickly infer the  environmental state but incurs large energetic losses in lean times.
Effective collective performance therefore requires a division of labor between a minority that forages even when rewards are unlikely and a majority that waits for favorable conditions. 

\begin{figure}[b!]
    \centering
    \includegraphics[width=.66\linewidth]{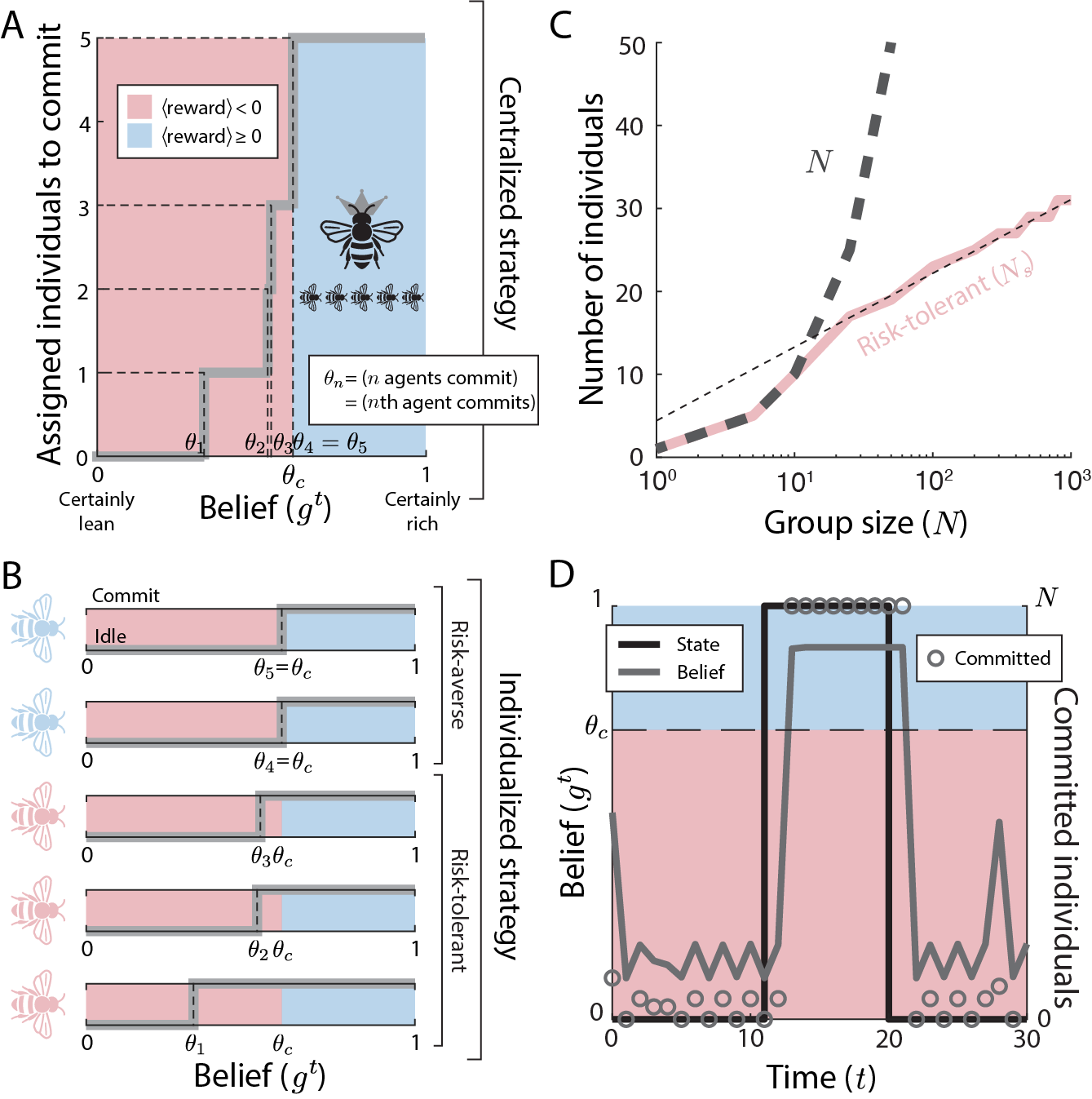}
    \caption{\textbf{Division of labor enables centralized and decentralized coordination.}
    \textbf{(A)}~In decentralized collectives, individuals have different foraging thresholds. Risk-averse individuals share a common threshold, $\theta_c,$ and forage only when the expected reward is positive. Risk-tolerant explorers have lower, heterogeneous thresholds and forage even when the expected reward is negative. 
    \textbf{(B)}~Under centralized control, a planner implementing the optimal policy, $\pi(g),$ deploys an increasing number of foragers as the belief increases. An appropriate distribution of individual thresholds allows a decentralized collective to match this allocation (compare panels A and B). 
    \textbf{(C)}~The optimal number of risk-tolerant explorers, $N_s,$ grows sublinearly with population size, $N$, so an increasingly small fraction of explorers suffices for efficient foraging. 
    \textbf{(D)}~The collective belief, $g^t,$ tracks environmental fluctuations because a subset continues exploring during lean periods, so the collective can rapidly re-engage when conditions improve.}
    \label{fig:opt}
\end{figure}

\begin{figure*}[b!]
    \centering
    \includegraphics[width=.99\linewidth]{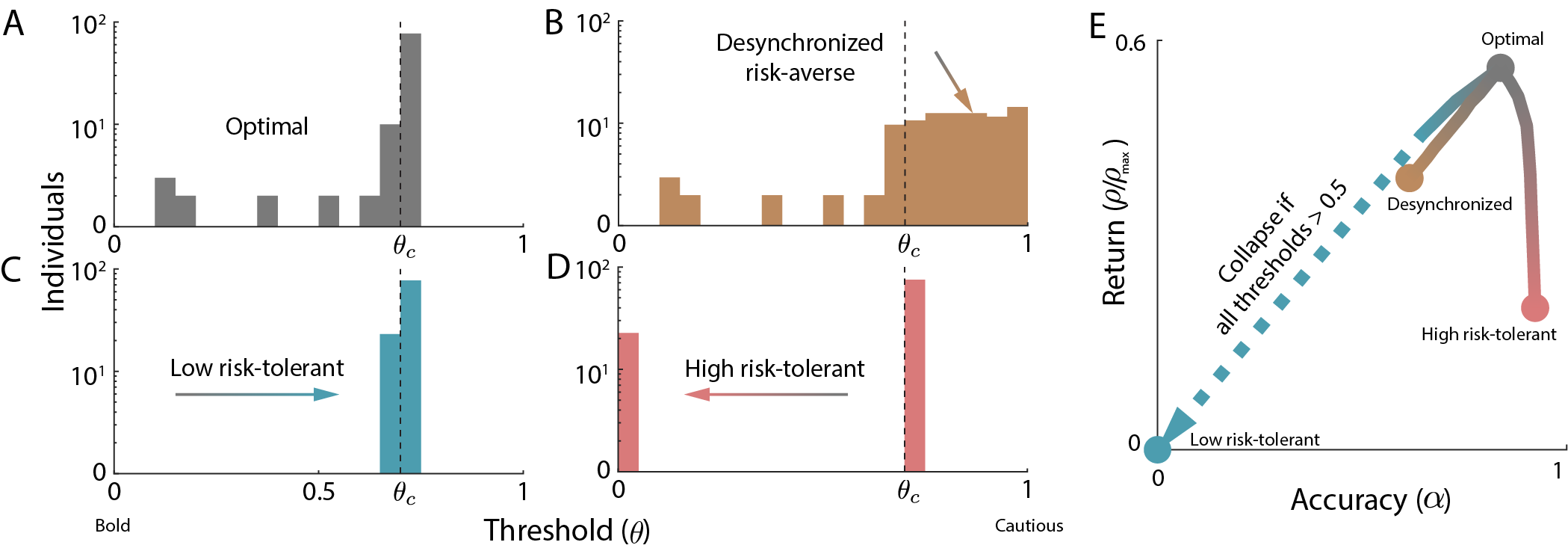}
    \caption{\textbf{Effects of perturbing the distribution of decision thresholds.}
    \textbf{(A)}~Optimal threshold distribution for a collective of $N=100$ individuals.
    \textbf{(B-D)}~Perturbed strategy distributions:
    \textbf{(B)}~increasing the spread of thresholds in the risk-averse majority;
    \textbf{(C)}~risk-tolerant individuals sharing a high threshold;
    \textbf{(D)}~risk-tolerant individuals sharing a low threshold.
    Colors indicate the different perturbation types.
    \textbf{(E)}~Normalized return vs. accuracy for the optimal strategy and its perturbations shows that returns are maximized at intermediate accuracy (gray dot; panel~A), reflecting weakened coordinated exploitation under desynchronization (brown–gray curve; panel~B), improved accuracy at increased cost under moderate heterogeneity (pink–gray curve; panel~C), and degraded returns under excessive risk tolerance (blue–gray curve; panel~D). Returns are normalized by the maximum possible return $\rho_{\max}$, defined as the return obtained when the entire collective forages only in the high-reward state.
    }
    \label{fig:perturb}
\end{figure*}

With an appropriate distribution of individual decision thresholds, a decentralized collective can achieve the same expected returns as an optimally controlled group. We show this by constructing a threshold distribution that reproduces the number of foragers deployed under the optimal centralized policy.
We do not suggest that collectives compute or implement a centralized solution; rather, this construction demonstrates that simple local threshold rules can recover centralized allocations, allowing us to analyze how ecological parameters shape the structure and robustness of the resulting strategy.
The optimal strategy of a central planner can be represented by a function, $\pi(g),$ that maps the collective belief about the state of the environment, $g,$ to the number of individuals sent out to forage, $n = \pi(g)$ (Fig.~\ref{fig:opt}A). 
As the belief that the environment is in a high-reward state increases, the planner commits more foragers \cite{Puterman1994,Topkis1998}. 
The optimal policy, $\pi(g),$ is non-decreasing in the belief $g$, since higher expected rewards never justify deploying fewer foragers. We can therefore define a sequence of belief thresholds, $\theta_n = \inf\{ \theta \in [0,1] : \pi(\theta) \geq n\}$ for $n = 1,2,\cdots,N$, where $N$ is the group size.  
Each threshold, $\theta_n,$ is the smallest belief value at which sending out at least $n$ foragers is optimal. Multiple thresholds may coincide, and in large groups the majority of thresholds equal the critical belief $\theta_c$ (Fig.~\ref{fig:perturb}) at which the expected foraging return is zero.

The strategy of a decentralized collective is fully determined by a set of private thresholds. Hence, a collective of size $N$ whose individuals have thresholds $\{\theta_n\}_{n=1}^N$ will send out exactly $n$ foragers when the belief satisfies $\theta_n \leq  g  < \theta_{n+1}$, with the convention $\theta_{N+1} = 1$ (Fig.~\ref{fig:opt}B). 
This choice of thresholds thus results in a number of foragers equal to that dictated by the optimal policy, $\pi(g)$ for any belief $g$. Importantly, individual decisions are based on a shared belief paired with private thresholds, rather than on orders from a central planner.
The resulting distribution of foraging strategies could arise from evolved or learned differences in risk sensitivity~\cite{Seeley1983,Dreller1998,Liang2012}, and is consistent with classic response-threshold models of division of labor~\cite{Bonabeau1996,Beshers2001}.

Collectives governed by a central decision-maker are inherently vulnerable to the loss or failure of this planner. In contrast, decentralized collectives with heterogeneous individual strategies are more robust to decision noise and environmental perturbations, potentially promoting evolutionary stability~\cite{Jandt2013,Couzin2005}.
As long as phenotypic differences across the population generate an appropriate distribution of risk tolerances, the collective can achieve optimal foraging performance~\cite{Seeley1983,Charbonneau2015}. 
Structured heterogeneity thus allows decentralized groups to match the allocations of a central planner while maintaining robust performance in dynamic environments. We next analyze the structure of the optimal population and ask how ecological pressures shape this distribution of strategies.

\subsection{Minority exploration and majority synchronization}

An optimal collective balances risk-tolerant individuals, who forage under uncertainty, with risk-averse individuals, who forage only when the expected reward is positive.
Empirical studies show a consistent pattern in many social species: a small fraction of individuals, such as ``scouts" in honeybee populations, persistently engages in exploration, while most colony members adopt safer strategies and exploit known resources~\cite{Seeley1983,Seeley2011,Beekman2001,Dornhaus2006}.
We show that the optimal decentralized  strategy yields a similar pattern. 
To do so, we use the critical threshold, $\theta_c,$ and classify individuals with decision thresholds below $\theta_c$ as risk-tolerant, and those with thresholds equal to or exceeding $\theta_c$ as risk-averse.

In small optimal collectives, all individuals are risk-tolerant, because each foraging outcome is highly informative and has a large impact on the  collective belief about the environmental state (Fig.~\ref{fig:opt}C). As group size increases, performance improves, but the optimal division of labor changes. The expected cost of foraging grows linearly with the number of active foragers, whereas  additional foraging outcomes provide redundant information and thus become less valuable.
A mathematical analysis shows that the optimal number of risk-tolerant individuals grows logarithmically with group size (dashed line in Fig.~\ref{fig:opt}C), reflecting diminishing informational returns from additional explorers as observations become increasingly redundant.
This sublinear scaling is robust across a wide range of environmental volatilities, foraging costs, and reward structures (SI Appendix, Fig.~S1).
Consequently, larger groups comprise a growing majority of risk-averse individuals, while the fraction of risk-tolerant explorers declines.
This prediction is consistent with empirical and theoretical studies showing that effective animal collectives rely on a small exploratory minority whose size grows more slowly than group size~\cite{Couzin2005,Cronin2014}.
Details of the asymptotic analysis underlying this result are provided in SI Appendix, Section 2.

While the decision thresholds of risk-tolerant individuals are distributed, all risk-averse individuals share a common threshold, $\theta_c$.
Indeed, the entire collective commits to foraging once the expected reward becomes positive (see SI Appendix and Fig.~S2).
This synchronization of foraging bouts is a consequence of the linear increase of total reward with the number of committed individuals, implying that full commitment is optimal when returns are expected to be favorable.
Thus, an optimal collective consists of a heterogeneous minority that maintains environmental awareness through continued sampling, and a synchronized majority that maximizes collective returns through coordinated exploitation once conditions are deemed favorable.

Optimal collective foraging strategies exhibit a characteristic cascade of activity in response to environmental changes (Fig.~\ref{fig:opt}D). When conditions deteriorate, widespread foraging produces few successes, quickly shifting the collective belief toward low expected rewards.
Most individuals stop foraging, leaving a small subset of risk-tolerant explorers that continue sampling.
When conditions deteriorate severely, even limited exploration ceases to minimize costs. Nevertheless, in the absence of new information, collective belief gradually relaxes toward uncertainty, prompting renewed exploration. 
If some of these sparse foraging attempts are successful, the belief shifts towards high expected rewards, and more foragers are recruited. Once the collective belief crosses the threshold $\theta_c$, coordinated collective foraging is triggered again, and the process repeats.
Heterogeneity in decision thresholds enables the population to function as a distributed sensor, maintaining sensitivity to environmental changes while balancing the cost of exploration.

Risk-tolerant individuals thus act as functional leaders by providing information about the environment, rather than by imposing their choices on others. 
Their exploratory actions inform the more cautious majority, resulting in a form of passive leadership that emerges from threshold heterogeneity\cite{King2009,Strandburg-Peshkin2018,Jolles2020}, rather than social dominance or intentional coordination.
This dual-speed system, driven by the gradual discoveries of risk-takers coupled with rapid collective withdrawal after mass failures in foraging attempts, allows for efficient collective responses without explicit coordination
(Fig.~\ref{fig:opt}D and SI Appendix, Fig. S3).

\subsection{Uniform strategies degrade collective performance}

A division of labor based on diverse risk tolerances is therefore crucial for balancing the costs and benefits of exploration (Fig.~\ref{fig:perturb}A). We next compare the performance of such optimal collectives to those with suboptimal strategy compositions.
We quantify collective performance using two metrics (Fig.~\ref{fig:perturb}E). The first is the long-term average return, $\rho$, normalized by the maximum possible return $\rho_{\max} = (\gamma_+ -\lambda)/2$ achieved when all individuals forage only when conditions are favorable.
The second is accuracy, $\alpha$, defined as the time-averaged proportion of individuals committed to foraging when the environment is in the high-reward state.

When all thresholds are high, the population is overly cautious, since even relatively risk-tolerant individuals forage too infrequently to detect environmental improvements. As a result, the collective responds slowly when good times return, reducing long-term average reward (Fig.~\ref{fig:perturb}C and E). In the extreme case where all individuals require excessive certainty to forage ($\theta_n > 1/2$ for all $n$), unsuccessful foraging drives the collective belief below $1/2$, halting further exploration. In the absence of new observations, the belief relaxes back toward uncertainty ($g=1/2$). Since all decision thresholds exceed this value, the collective remains inactive indefinitely. 

Conversely, a collective with too many risk-takers fails in the opposite way: eager explorers can maintain high accuracy, but only by sustaining near-continuous sampling. The marginal information gain of these explorers is outweighed by the energetic cost of foraging, depressing long-term returns (Fig.~\ref{fig:perturb}D and E). In contrast, unstructured deviations from a shared risk-averse threshold primarily impair exploitation rather than information acquisition, desynchronizing the cautious majority and preventing coordinated surges of foraging when conditions are favorable (Fig.~\ref{fig:perturb}B and E). 

Optimal collectives thus exhibit functional specialization, with a synchronized risk-averse majority sharing threshold $\theta_c$ responsible for coordinated exploitation, and a heterogeneous risk-tolerant minority that sustains exploration through a broad distribution of thresholds in large groups (Fig.~\ref{fig:perturb} and SI Appendix, Fig. S4).
Together, these results reveal a collective exploration–exploitation trade-off: heterogeneity among risk-tolerant individuals sustains exploration under uncertainty, while synchronization of the risk-averse majority maximizes returns once conditions are favorable.
We next ask how environmental conditions shape the optimal balance between explorers and exploiters, and identify when ecological pressures favor heterogeneous group compositions over uniform ones.

\subsection{Ecological extremes suppress division of labor}

\begin{figure*}[b!]
    \centering
    \includegraphics[width=.99\linewidth]{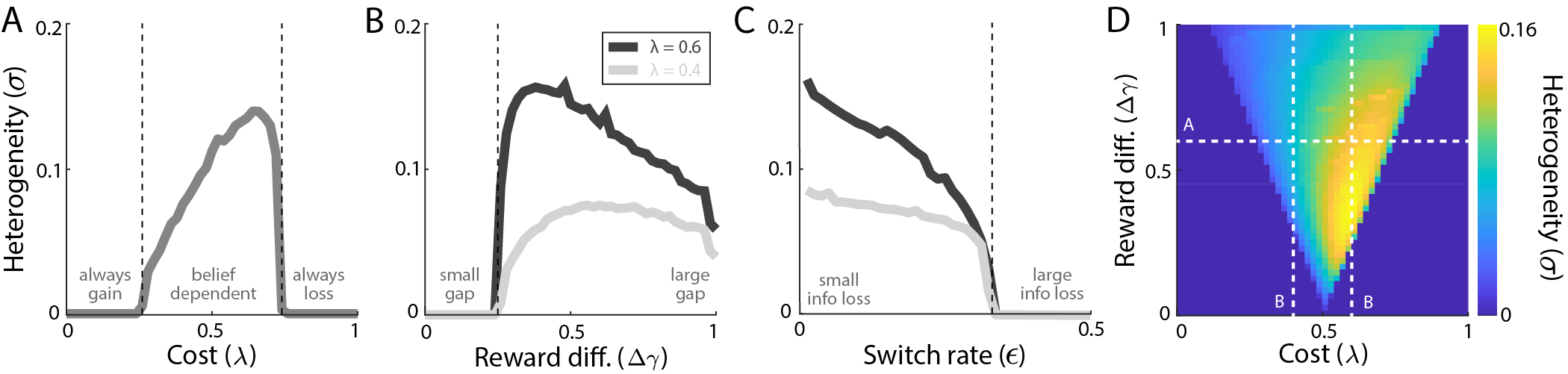}
    \caption{\textbf{Ecological modulation of collective heterogeneity.}
    \textbf{(A)}~Heterogeneity of optimal decision thresholds varies non-monotonically with commitment cost, $\lambda$, peaking at intermediate values where foraging decisions depend sensitively on belief; extreme costs lead to uniform inactivity (high cost) or uniform foraging (low cost).
    \textbf{(B)}~Heterogeneity also peaks at intermediate reward rate contrasts, $\Delta\gamma,$ (shown for two commitment costs): small changes in reward rate are both hard to detect and not worth detecting. Large changes are easily detected by a few foragers. Heterogeneity thus saturates in both limits.
    \textbf{(C)}~Rapid environmental switching suppresses heterogeneity by limiting achievable inference accuracy.
    \textbf{(D)}~Joint variation of cost and reward contrast reveals a wedge of intermediate ecological conditions in which division of labor and structured heterogeneity are optimal.}
    \label{fig:env}
\end{figure*}

We next ask how changes in environmental conditions impact the optimal division of labor between risk-tolerant and risk-averse individuals. In particular, we show that the cost of foraging, $\lambda$ (energetic expenditure and risk), the difference between expected reward during good and bad times, $\Delta \gamma = \gamma_+ - \gamma_-$; and environmental variability, $\epsilon$ (the probability of state transitions) strongly shape the structure of a well-calibrated collective. 
We quantify these effects using the standard deviation of the optimal threshold distribution as a measure of heterogeneity of strategies in the collective.

Group heterogeneity varies non-monotonically with foraging cost, peaking at intermediate values (Fig.~\ref{fig:env}A and D, SI Appendix, Fig.~S5).
When costs are low ($\theta_c(\lambda) < \epsilon$), resources are easily accessed and all individuals forage,  resulting in a homogeneous population of foragers.
Conversely, when costs are high ($1-\epsilon < \theta_c(\lambda)$), exploration becomes prohibitively expensive, and all individuals refrain from foraging. 
Between these extremes ($\epsilon \leq \theta_c(\lambda) \leq 1-\epsilon$), a maximal variability of risk-preference is attained. 
In this intermediate regime, moderate increases in foraging cost selectively suppress marginal exploration by cautious individuals, while favoring continued sampling by a smaller risk-tolerant subset, thereby amplifying the diversity of thresholds used (wedge-shaped region of high heterogeneity in Fig.~\ref{fig:env}D).

Group heterogeneity also varies non-monotonically with the difference between the expected reward rate in the two environmental states, $\Delta \gamma = \gamma_+ - \gamma_-$ (Fig.~\ref{fig:env}B and D, SI Appendix, Fig.~S5).
When this difference is small ($\theta_c(\Delta \gamma) < \epsilon$ or $\theta_c(\Delta \gamma) > 1-\epsilon$), the environment is nearly deterministic, and the population follows a homogeneous strategy. Again, all individuals forage or remain idle depending on whether the average reward exceeds the commitment cost.
As the difference increases, distinguishing favorable from unfavorable conditions becomes more valuable, as early detection yields large gains in expected return. This leads to the emergence of risk-tolerant individuals and greater strategic heterogeneity.
At extreme contrasts in reward, transitions to a favorable state are easily inferred without extensive exploration, because even sparse successes strongly shift belief, reducing the benefits of maintaining many risk-tolerant individuals in the population.
Heterogeneity is amplified when commitment costs are high, as risk-averse individuals require greater certainty  to forage, increasing the informational value of continued sampling by a small exploratory minority.

Rapid environmental changes suppress heterogeneity, favoring homogeneous risk aversion over exploratory strategies (Fig.~\ref{fig:env}C).
When environments fluctuate rapidly, maintaining risk-tolerant individuals becomes counterproductive as information gained through exploration quickly becomes obsolete.
In this regime, the timescale of environmental changes is shorter than the time needed to coordinate a response.
As a result, fast-changing environments favor conservative strategies that minimize exploration costs, leading to uniform populations of risk-averse individuals.
In contrast, when environmental changes are slow and predictable relative to the timescale of exploration and response, information accumulates and remains reliable over long periods, so the group can profit from exploratory foraging in future decisions. Under these conditions, risk-tolerant individuals can repeatedly detect  state changes, stabilizing collective belief and enabling timely collective commitment, favoring heterogeneous strategies.

\section*{Discussion}

We have shown that leaderless collectives can achieve optimal foraging performance through a division of labor based on heterogeneous risk tolerances. 
A small fraction of risk-tolerant individuals, whose number increases logarithmically with group size, maintains environmental awareness by exploring even when expected rewards are negative, while a synchronized risk-averse majority maximizes collective gains by committing only when conditions are favorable.
This asymmetric structure reflects fundamental principles of collective information gathering: as individuals sample independently, the information value of additional observations diminishes due to redundancy, whereas exploration costs accumulate linearly with the number of explorers, so only a small fraction is needed for accurate environmental tracking in large groups.

Optimal collective foraging exhibits characteristic decision cascades in response to environmental changes. Minority-driven exploration triggers rapid group-wide recruitment when conditions improve, enabling swift responses to opportunities while limiting costly false alarms. Exploration loses its value under extreme ecological pressures or in rapidly changing environments where information becomes obsolete before it can guide actions. This causes collectives to favor uniform risk aversion over strategic diversity.
Hence, heterogeneity in decision thresholds is most pronounced under intermediate foraging costs and environmental contrasts (Fig.~\ref{fig:env}).  

Risk-tolerant individuals function as {\em informative leaders} in decentralized groups, guiding collective behavior by maintaining environmental awareness during periods of uncertainty~\cite{Strandburg-Peshkin2018,Couzin2005,King2009}. Whether such leadership is persistent or transient depends on its origin. Stable phenotypic variation in energetic state or temperament can produce consistent leaders~\cite{Biro2008,King2009,Jolles2017}, whereas probabilistic decision-making by otherwise homogeneous individuals allows leadership to rotate while preserving near-optimal collective performance~\cite{Couzin2011,Conradt2005,Jolles2020} (See SI Appendix, Section 4 for proof).

While comparable divisions of labor can be imposed by centralized or dominance-based control in some groups~\cite{Conradt2005,Mech1999}, we focused on how informational leadership can emerge in leaderless collectives via heterogeneity in exploration preference under uncertainty~\cite{Couzin2009,Hein2015}.
A minority that samples when payoffs are ambiguous can distribute group sensing to regulate majority engagement, consistent with broader views of collective computation and information flow in animal groups~\cite{Sumpter2008}.
This perspective suggests that if exploratory roles reflect stable phenotypic differences, individuals should show cross-context consistency in risk-taking and influence~\cite{Biro2008,Jolles2017,King2018}; if instead exploration is due to stochastic choices among otherwise similar individuals, leadership should rotate without loss of group-level performance~\cite{Couzin2011,Mann2018}. 
Correspondingly, altering exploration costs or environmental persistence is expected to shift the degree and stability of leadership and the strength of division of labor~\cite{Lima1990,Dall2005}. These predictions are experimentally accessible in social insects and other tractable systems where ecological timescales and costs can be manipulated and collective responses quantified~\cite{Seeley1994,Davidson2016,Dornhaus2006}.

A sublinearly increasing number of information leaders appears in diverse collective decision-making systems: informed minorities can effectively guide na\"{i}ve majorities in moving animal groups~\cite{Couzin2005}, a small fraction of hasty explorers can optimize group decision efficiency when observed by a large, deliberate majority~\cite{Karamched2020}, and a few scouts can reliably inform nest-site choices in social insects~\cite{Cronin2014}.
Although leadership by risk-tolerant minorities has been studied experimentally in groups of fixed size~\cite{Reebs2000,Beekman2006,Dyer2008}, how the number of leaders should scale with group size and how this scaling emerges from individual decision rules remains largely unexplored.

The role of heterogeneity in group composition has been extensively studied using threshold-based models of division of labor, which typically assume symmetric, normally distributed response thresholds across individuals~\cite{Bonabeau1996,Bonabeau1998,Beshers2001,OSheaWheller2017,Ulrich2021}. However, the structure of threshold distributions itself has been largely overlooked. Skewness alone can generate complex relationships between group composition and collective performance. Division of labor in social insects can emerge from multiple interacting factors, beyond threshold variation~\cite{Ulrich2021}, yet the contribution of the distribution structure remains unexplored. 
We hypothesize that natural groups may evolve skewed threshold compositions  to optimize collective task performance.

One limitation of our model is the assumption of perfect information sharing as all individuals instantaneously observe collective foraging outcomes. In many natural systems not all potential foragers observe every returning individual, and information is transmitted locally or indirectly~\cite{Sumpter2005,Sumpter2008,Karamched2020}. Such local transmission can generate modular organization, in which distinct subpopulations have unequal access to exploratory returns. Even when information is available, individuals may differ in perceptual or decision noise (e.g., noisy belief updates or variable thresholds)~\cite{Bogacz2006}, resulting in heterogeneous beliefs and stochastic actions.  We expect the qualitative division-of-labor strategy identified here to persist with such noise.
Under such conditions, individuals may rely not only on explicit signals but also on implicit social information inferred from the behavior of others~\cite{Mann2018,Mann2020,Suganuma2025}.
How partial information sharing and social inference reshape the optimal division of labor remains an open question~\cite{Beshers2001}.
We hypothesize that the core organizational structure identified here -- a  minority of explorers along with a synchronized majority -- remains robust under imperfect communication, although achieving comparable performance may require a larger exploratory cohort to compensate for information loss.

Our results also raise  questions about how finite energetic reserves constrain collective decision-making. While our model implicitly assumes unbounded accumulation of returns, biological systems operate under both lower and upper energetic limits.
Near starvation, collectives may be forced into riskier strategies to avoid extinction. How such transitions are distributed across heterogeneous groups remains unclear.
If risk heterogeneity reflects hunger-driven phenotypic variation, individuals with depleted reserves may become disproportionately risk-tolerant \cite{Theraulaz1998}, amplifying the influence of an informed minority during critical periods~\cite{King2009}.
Conversely, when storage approaches saturation, continued foraging becomes unnecessary, potentially suppressing exploration and diminishing the role of exploratory individuals altogether.
Understanding how energetic constraints reshape the emergence, persistence, and influence of informed minorities is therefore essential for linking individual state, division of labor, and collective responsiveness of natural systems.

In human society, the parallel to the risk-tolerant explorer is thus not the state-sponsored adventurer like Zheng He or Magellan, but the self-motivated innovator: the garage inventor, the citizen scientist, the open-source builder, the high-risk entrepreneur, and the avant-garde artist. These individuals are driven by curiosity and passion and have lower thresholds for action than others. They often bear substantial personal costs, yet when they succeed, they spark movements that drive collective behavior at scale.

\section{Methods}
\subsection*{Bellman equation and numerical solution}
To determine the optimal collective foraging strategy under a centralized planner, we formulate the  problem as a Markov decision process in which the system state is given by the collective belief about the environmental state. The goal is to find the strategy that maximizes the long-term average collective return. Using dynamic programming, we derive the Bellman equation characterizing the optimal policy, $\pi(g),$ that maps belief states $g$ to the number of agents committed to foraging. The derivation of the Bellman equation is provided in SI Appendix, Section 1.

We compute the optimal policy numerically by discretizing the belief space on the interval $[0,1]$ using a uniform grid of $200$ points and solving the discretized Bellman equation via fixed-point iteration. Iteration is continued until the value function converges to tolerance $10^{-6}$. The resulting policy serves as a benchmark for collective performance and as a reference for constructing decentralized strategies.

\subsection*{Asymptotic analysis and large deviation theory}
To understand how the optimal collective strategy, $\pi(g),$ changes with group size $N$, we perform asymptotic analysis of the Bellman equation in the limit of large $N$. This analysis characterizes the leading-order behavior of the optimal policy and shows that only a sublinear number of agents explore (forage) under unfavorable conditions, \emph{i.e.}, when $g < \theta_c$. In our model setup, this scaling is  logarithmic in $N$.

Rare events associated with future rewards and belief transitions are characterized using large deviation theory. These  estimates allow us to quantify the contribution of the current number of committed agents to long-term returns and to derive bounds on the optimal allocation of exploratory effort for large groups. Technical assumptions and full derivations are presented in SI Appendix, Section 2.

\subsection*{Perturbations of the optimal policy}
To perturb the optimal policy in Fig.~\ref{fig:perturb}, we construct  strategies by interpolating between the optimal policy $\pi(g)$ and three extreme reference policies $\pi_\text{ext}(g)$ corresponding to uniformly bold, uniformly cautious, and desynchronized cautious strategies. More precisely, the perturbed policy is given by $\pi_\lambda(g) = \lfloor(1-\lambda) \pi(g) + \lambda \pi_\text{ext}(g)\rfloor$ for $\lambda \in [0,1]$, where $\lfloor z\rfloor$ denotes the largest integer less than or equal to $z$. These interpolations systematically alter the degree of heterogeneity and synchronization in foraging behavior. Performance under perturbed policies is evaluated using the same metrics as for the optimal strategy (return and accuracy).

\subsection*{Model parameters}
All figures in the main text are generated using a baseline parameter set with group size $N = 100$, reward probabilities $\gamma_\pm = 0.5 \pm 0.3$, switch probability $\epsilon = 0.1$, and commitment cost $\lambda = 0.62$ (and thus $\theta_c = (\lambda-\gamma_-)/(\gamma_+ - \gamma_-) = 0.7$) chosen to represent typical model behavior unless otherwise specified. In Fig.~\ref{fig:perturb}, we vary the interpolation parameter while holding other parameters fixed at baseline values. In Fig.~\ref{fig:env}, we choose $\gamma_\pm = (1 \pm \Delta \gamma)/2$.

\subsection*{Monte Carlo simulations}
To compute performance metrics for given model parameters, we employed Monte Carlo simulations of the stochastic foraging process. For each parameter set and (perturbed) policy, we simulate the collective dynamics over long time horizons to estimate the performance metrics. Each data point in Fig.~\ref{fig:perturb} is computed from $10^3$ independent realizations, each simulated for $2\times10^4$ time steps.

\bibliographystyle{unsrtnat}
\bibliography{reference.bib}

@article{Couzin2009,
  title = {Collective cognition in animal groups},
  volume = {13},
  ISSN = {1364-6613},
  url = {http://dx.doi.org/10.1016/j.tics.2008.10.002},
  DOI = {10.1016/j.tics.2008.10.002},
  number = {1},
  journal = {Trends in Cognitive Sciences},
  publisher = {Elsevier BV},
  author = {Couzin,  Iain D.},
  year = {2009},
  month = jan,
  pages = {36–43}
}

@article{Conradt2003,
  title = {Group decision-making in animals},
  volume = {421},
  ISSN = {1476-4687},
  url = {http://dx.doi.org/10.1038/nature01294},
  DOI = {10.1038/nature01294},
  number = {6919},
  journal = {Nature},
  publisher = {Springer Science and Business Media LLC},
  author = {Conradt,  L. and Roper,  T. J.},
  year = {2003},
  month = jan,
  pages = {155–158}
}

@book{Stephens1986,
  title={Foraging theory},
  author={Stephens, David W and Krebs, John R},
  volume={6},
  year={1986},
  publisher={Princeton university press}
}

@article{Lima1990,
  title = {Behavioral decisions made under the risk of predation: a review and prospectus},
  volume = {68},
  ISSN = {1480-3283},
  url = {http://dx.doi.org/10.1139/z90-092},
  DOI = {10.1139/z90-092},
  number = {4},
  journal = {Canadian Journal of Zoology},
  publisher = {Canadian Science Publishing},
  author = {Lima,  Steven L. and Dill,  Lawrence M.},
  year = {1990},
  month = apr,
  pages = {619–640}
}

@article{Wolf1989,
  title = {Extra Loads and Foraging Life Span in Honeybee Workers},
  volume = {58},
  ISSN = {0021-8790},
  url = {http://dx.doi.org/10.2307/5134},
  DOI = {10.2307/5134},
  number = {3},
  journal = {The Journal of Animal Ecology},
  publisher = {JSTOR},
  author = {Wolf,  Thomas J. and Schmid-Hempel,  Paul},
  year = {1989},
  month = oct,
  pages = {943}
}

@article{King2018,
  title = {Re-wilding Collective Behaviour: An Ecological Perspective},
  volume = {33},
  ISSN = {0169-5347},
  url = {http://dx.doi.org/10.1016/j.tree.2018.03.004},
  DOI = {10.1016/j.tree.2018.03.004},
  number = {5},
  journal = {Trends in Ecology \& Evolution},
  publisher = {Elsevier BV},
  author = {King,  Andrew J. and Fehlmann,  Gaëlle and Biro,  Dora and Ward,  Ashley J. and F\"{u}rtbauer,  Ines},
  year = {2018},
  month = may,
  pages = {347–357}
}

@article{Mech1999,
  title = {Alpha status,  dominance,  and division of labor in wolf packs},
  volume = {77},
  ISSN = {1480-3283},
  url = {http://dx.doi.org/10.1139/z99-099},
  DOI = {10.1139/z99-099},
  number = {8},
  journal = {Canadian Journal of Zoology},
  publisher = {Canadian Science Publishing},
  author = {Mech,  L David},
  year = {1999},
  month = nov,
  pages = {1196–1203}
}

@article{Seeley2004,
  title = {Quorum sensing during nest-site selection by honeybee swarms},
  volume = {56},
  ISSN = {1432-0762},
  url = {http://dx.doi.org/10.1007/s00265-004-0814-5},
  DOI = {10.1007/s00265-004-0814-5},
  number = {6},
  journal = {Behavioral Ecology and Sociobiology},
  publisher = {Springer Science and Business Media LLC},
  author = {Seeley,  Thomas D. and Visscher,  P. Kirk},
  year = {2004},
  month = jul,
  pages = {594–601}
}

@article{Hein2015,
  title={The evolution of distributed sensing and collective computation in animal populations},
  author={Hein, Andrew M and Rosenthal, Sara Brin and Hagstrom, George I and Berdahl, Andrew and Torney, Colin J and Couzin, Iain D},
  journal={Elife},
  volume={4},
  pages={e10955},
  year={2015},
  publisher={eLife Sciences Publications, Ltd},
  url = {http://dx.doi.org/10.7554/eLife.10955}
}

@article{Dornhaus2006,
  title = {Colony size affects collective decision-making in the ant Temnothorax albipennis},
  volume = {53},
  ISSN = {1420-9098},
  url = {http://dx.doi.org/10.1007/s00040-006-0887-4},
  DOI = {10.1007/s00040-006-0887-4},
  number = {4},
  journal = {Insectes Sociaux},
  publisher = {Springer Science and Business Media LLC},
  author = {Dornhaus,  A. and Franks,  N. R.},
  year = {2006},
  month = nov,
  pages = {420–427}
}

@article{Cronin2014,
  title={Ants work harder during consensus decision-making in small groups},
  author={Cronin, Adam L and Stumpe, Martin C},
  journal={Journal of the Royal Society Interface},
  volume={11},
  number={98},
  pages={20140641},
  year={2014},
  publisher={The Royal Society}
}

@inbook{Bressloff2021,
  title = {Probability theory and martingales},
  ISBN = {9783030725150},
  ISSN = {2196-9973},
  url = {http://dx.doi.org/10.1007/978-3-030-72515-0_9},
  DOI = {10.1007/978-3-030-72515-0_9},
  booktitle = {Stochastic Processes in Cell Biology},
  publisher = {Springer International Publishing},
  author = {Bressloff,  Paul C.},
  year = {2021},
  pages = {681–703}
}

@book{Dembo2010,
  title = {Large Deviations Techniques and Applications},
  ISBN = {9783642033117},
  ISSN = {0172-4568},
  url = {http://dx.doi.org/10.1007/978-3-642-03311-7},
  DOI = {10.1007/978-3-642-03311-7},
  journal = {Stochastic Modelling and Applied Probability},
  publisher = {Springer Berlin Heidelberg},
  author = {Dembo,  Amir and Zeitouni,  Ofer},
  year = {2010}
}

@article{Beekman2001,
  title = {Phase transition between disordered and ordered foraging in Pharaoh’s ants},
  volume = {98},
  ISSN = {1091-6490},
  url = {http://dx.doi.org/10.1073/pnas.161285298},
  DOI = {10.1073/pnas.161285298},
  number = {17},
  journal = {Proceedings of the National Academy of Sciences},
  publisher = {Proceedings of the National Academy of Sciences},
  author = {Beekman,  Madeleine and Sumpter,  David J. T. and Ratnieks,  Francis L. W.},
  year = {2001},
  month = aug,
  pages = {9703–9706}
}

@article{Jandt2013,
  title = {Behavioural syndromes and social insects: personality at multiple levels},
  volume = {89},
  ISSN = {1469-185X},
  url = {http://dx.doi.org/10.1111/brv.12042},
  DOI = {10.1111/brv.12042},
  number = {1},
  journal = {Biological Reviews},
  publisher = {Wiley},
  author = {Jandt,  Jennifer M. and Bengston,  Sarah and Pinter‐Wollman,  Noa and Pruitt,  Jonathan N. and Raine,  Nigel E. and Dornhaus,  Anna and Sih,  Andrew},
  year = {2013},
  month = may,
  pages = {48–67}
}

@article{Seeley1983,
  title = {Division of labor between scouts and recruits in honeybee foraging},
  volume = {12},
  ISSN = {1432-0762},
  url = {http://dx.doi.org/10.1007/BF00290778},
  DOI = {10.1007/bf00290778},
  number = {3},
  journal = {Behavioral Ecology and Sociobiology},
  publisher = {Springer Science and Business Media LLC},
  author = {Seeley,  Thomas D.},
  year = {1983},
  month = jun,
  pages = {253–259}
}

@article{Davidson2016,
  title = {Effect of Interactions between Harvester Ants on Forager Decisions},
  volume = {4},
  ISSN = {2296-701X},
  url = {http://dx.doi.org/10.3389/fevo.2016.00115},
  DOI = {10.3389/fevo.2016.00115},
  journal = {Frontiers in Ecology and Evolution},
  publisher = {Frontiers Media SA},
  author = {Davidson,  Jacob D. and Arauco-Aliaga,  Roxana P. and Crow,  Sam and Gordon,  Deborah M. and Goldman,  Mark S.},
  year = {2016},
  month = oct 
}

@article{Seeley1994,
  title = {Honey bee foragers as sensory units of their colonies},
  volume = {34},
  ISSN = {1432-0762},
  url = {http://dx.doi.org/10.1007/BF00175458},
  DOI = {10.1007/bf00175458},
  number = {1},
  journal = {Behavioral Ecology and Sociobiology},
  publisher = {Springer Science and Business Media LLC},
  author = {Seeley,  Thomas D.},
  year = {1994},
  month = jan,
  pages = {51–62}
}

@article{BlumMoyse2025,
  title = {Social Patch Foraging Theory in an Egalitarian Group},
  volume = {3},
  ISSN = {2835-8279},
  url = {http://dx.doi.org/10.1103/mqxr-1nyv},
  DOI = {10.1103/mqxr-1nyv},
  number = {3},
  journal = {PRX Life},
  publisher = {American Physical Society (APS)},
  author = {Blum Moyse,  Lisa and El Hady,  Ahmed},
  year = {2025},
  month = jul 
}

@article{Jolles2020,
  title = {The Role of Individual Heterogeneity in Collective Animal Behaviour},
  volume = {35},
  ISSN = {0169-5347},
  url = {http://dx.doi.org/10.1016/j.tree.2019.11.001},
  DOI = {10.1016/j.tree.2019.11.001},
  number = {3},
  journal = {Trends in Ecology \& Evolution},
  publisher = {Elsevier BV},
  author = {Jolles,  Jolle W. and King,  Andrew J. and Killen,  Shaun S.},
  year = {2020},
  month = mar,
  pages = {278–291}
}

@article{King2009,
  title = {The Origins and Evolution of Leadership},
  volume = {19},
  ISSN = {0960-9822},
  url = {http://dx.doi.org/10.1016/j.cub.2009.07.027},
  DOI = {10.1016/j.cub.2009.07.027},
  number = {19},
  journal = {Current Biology},
  publisher = {Elsevier BV},
  author = {King,  Andrew J. and Johnson,  Dominic D.P. and Van Vugt,  Mark},
  year = {2009},
  month = oct,
  pages = {R911–R916}
}

@article{Couzin2005,
  title = {Effective leadership and decision-making in animal groups on the move},
  volume = {433},
  ISSN = {1476-4687},
  url = {http://dx.doi.org/10.1038/nature03236},
  DOI = {10.1038/nature03236},
  number = {7025},
  journal = {Nature},
  publisher = {Springer Science and Business Media LLC},
  author = {Couzin,  Iain D. and Krause,  Jens and Franks,  Nigel R. and Levin,  Simon A.},
  year = {2005},
  month = feb,
  pages = {513–516}
}

@article{Strandburg-Peshkin2018,
  title = {Inferring influence and leadership in moving animal groups},
  volume = {373},
  ISSN = {1471-2970},
  url = {http://dx.doi.org/10.1098/rstb.2017.0006},
  DOI = {10.1098/rstb.2017.0006},
  number = {1746},
  journal = {Philosophical Transactions of the Royal Society B: Biological Sciences},
  publisher = {The Royal Society},
  author = {Strandburg-Peshkin,  Ariana and Papageorgiou,  Danai and Crofoot,  Margaret C. and Farine,  Damien R.},
  year = {2018},
  month = mar,
  pages = {20170006}
}

@book{degroot2005optimal,
  title={Optimal Statistical Decisions},
  author={DeGroot, M.H.},
  isbn={9780471726142},
  series={Wiley Classics Library},
  url={https://books.google.com/books?id=dtVieJ245z0C},
  year={2005},
  publisher={Wiley}
}

@article{Karamched2020,
  title={Heterogeneity improves speed and accuracy in social networks},
  author={Karamched, Bhargav and Stickler, Megan and Ott, William and Lindner, Benjamin and Kilpatrick, Zachary P and Josi{\'c}, Kre{\v{s}}imir},
  journal={Physical review letters},
  volume={125},
  number={21},
  pages={218302},
  year={2020},
  publisher={APS}
}

@article{Beekman2006,
  title = {How does an informed minority of scouts guide a honeybee swarm as it flies to its new home?},
  volume = {71},
  ISSN = {0003-3472},
  url = {http://dx.doi.org/10.1016/j.anbehav.2005.04.009},
  DOI = {10.1016/j.anbehav.2005.04.009},
  number = {1},
  journal = {Animal Behaviour},
  publisher = {Elsevier BV},
  author = {Beekman,  Madeleine and Fathke,  Robert L. and Seeley,  Thomas D.},
  year = {2006},
  month = jan,
  pages = {161–171}
}

@article{Reebs2000,
  title = {Can a minority of informed leaders determine the foraging movements of a fish shoal?},
  volume = {59},
  ISSN = {0003-3472},
  url = {http://dx.doi.org/10.1006/anbe.1999.1314},
  DOI = {10.1006/anbe.1999.1314},
  number = {2},
  journal = {Animal Behaviour},
  publisher = {Elsevier BV},
  author = {Reebs,  Stephan G.},
  year = {2000},
  month = feb,
  pages = {403–409}
}

@article{Dyer2008,
  title = {Leadership,  consensus decision making and collective behaviour in humans},
  volume = {364},
  ISSN = {1471-2970},
  url = {http://dx.doi.org/10.1098/rstb.2008.0233},
  DOI = {10.1098/rstb.2008.0233},
  number = {1518},
  journal = {Philosophical Transactions of the Royal Society B: Biological Sciences},
  publisher = {The Royal Society},
  author = {Dyer,  John R.G and Johansson,  Anders and Helbing,  Dirk and Couzin,  Iain D and Krause,  Jens},
  year = {2008},
  month = dec,
  pages = {781–789}
}

@article{Franks2003,
  title = {Speed versus accuracy in collective decision making},
  volume = {270},
  ISSN = {1471-2954},
  url = {http://dx.doi.org/10.1098/rspb.2003.2527},
  DOI = {10.1098/rspb.2003.2527},
  number = {1532},
  journal = {Proceedings of the Royal Society of London. Series B: Biological Sciences},
  publisher = {The Royal Society},
  author = {Franks,  Nigel R. and Dornhaus,  Anna and Fitzsimmons,  Jon P. and Stevens,  Martin},
  year = {2003},
  month = dec,
  pages = {2457–2463}
}

@book{Seeley2011,
  title={Honeybee democracy},
  author={Seeley, Thomas D},
  year={2011},
  publisher={Princeton University Press}
}

@article{Couzin2011,
  title = {Uninformed Individuals Promote Democratic Consensus in Animal Groups},
  volume = {334},
  ISSN = {1095-9203},
  url = {http://dx.doi.org/10.1126/science.1210280},
  DOI = {10.1126/science.1210280},
  number = {6062},
  journal = {Science},
  publisher = {American Association for the Advancement of Science (AAAS)},
  author = {Couzin,  Iain D. and Ioannou,  Christos C. and Demirel,  G\"{u}ven and Gross,  Thilo and Torney,  Colin J. and Hartnett,  Andrew and Conradt,  Larissa and Levin,  Simon A. and Leonard,  Naomi E.},
  year = {2011},
  month = dec,
  pages = {1578–1580}
}

@article{Kilpatrick2021,
  title = {Uncertainty drives deviations in normative foraging decision strategies},
  volume = {18},
  ISSN = {1742-5662},
  url = {http://dx.doi.org/10.1098/rsif.2021.0337},
  DOI = {10.1098/rsif.2021.0337},
  number = {180},
  journal = {Journal of The Royal Society Interface},
  publisher = {The Royal Society},
  author = {Kilpatrick,  Zachary P. and Davidson,  Jacob D. and El Hady,  Ahmed},
  year = {2021},
  month = jul,
  pages = {20210337}
}

@article{Masuda2015,
  title = {Computational model of collective nest selection by ants with heterogeneous acceptance thresholds},
  volume = {2},
  ISSN = {2054-5703},
  url = {http://dx.doi.org/10.1098/rsos.140533},
  DOI = {10.1098/rsos.140533},
  number = {6},
  journal = {Royal Society Open Science},
  publisher = {The Royal Society},
  author = {Masuda,  Naoki and O’shea-Wheller,  Thomas A. and Doran,  Carolina and Franks,  Nigel R.},
  year = {2015},
  month = jun,
  pages = {140533}
}

@article{OSheaWheller2017,
  title = {Variability in individual assessment behaviour and its implications for collective decision-making},
  volume = {284},
  ISSN = {1471-2954},
  url = {http://dx.doi.org/10.1098/rspb.2016.2237},
  DOI = {10.1098/rspb.2016.2237},
  number = {1848},
  journal = {Proceedings of the Royal Society B: Biological Sciences},
  publisher = {The Royal Society},
  author = {O’Shea-Wheller,  Thomas A. and Masuda,  Naoki and Sendova-Franks,  Ana B. and Franks,  Nigel R.},
  year = {2017},
  month = feb,
  pages = {20162237}
}

@article{Robinson2011,
  title = {A Simple Threshold Rule Is Sufficient to Explain Sophisticated Collective Decision-Making},
  volume = {6},
  ISSN = {1932-6203},
  url = {http://dx.doi.org/10.1371/journal.pone.0019981},
  DOI = {10.1371/journal.pone.0019981},
  number = {5},
  journal = {PLoS ONE},
  publisher = {Public Library of Science (PLoS)},
  author = {Robinson,  Elva J. H. and Franks,  Nigel R. and Ellis,  Samuel and Okuda,  Saki and Marshall,  James A. R.},
  editor = {Adler,  Frederick R.},
  year = {2011},
  month = may,
  pages = {e19981}
}

@book{Puterman1994,
  title = {Markov Decision Processes: Discrete Stochastic Dynamic Programming},
  ISBN = {9780470316887},
  ISSN = {1940-6347},
  url = {http://dx.doi.org/10.1002/9780470316887},
  DOI = {10.1002/9780470316887},
  journal = {Wiley Series in Probability and Statistics},
  publisher = {Wiley},
  author = {Puterman,  Martin L.},
  year = {1994},
  month = apr 
}

@article{Mahadevan1996,
  title = {Average reward reinforcement learning: Foundations,  algorithms,  and empirical results},
  volume = {22},
  ISSN = {1573-0565},
  url = {http://dx.doi.org/10.1007/BF00114727},
  DOI = {10.1007/bf00114727},
  number = {1–3},
  journal = {Machine Learning},
  publisher = {Springer Science and Business Media LLC},
  author = {Mahadevan,  Sridhar},
  year = {1996},
  pages = {159–195}
}

@article{Bellman1954,
  title={The theory of dynamic programming},
  author={Bellman, Richard},
  journal={Bulletin of the American Mathematical Society},
  volume={60},
  number={6},
  pages={503--515},
  year={1954}
}

@book{Frisch1993,
  title = {The Dance Language and Orientation of Bees},
  ISBN = {9780674418769},
  url = {http://dx.doi.org/10.4159/harvard.9780674418776},
  DOI = {10.4159/harvard.9780674418776},
  publisher = {Harvard University Press},
  author = {Frisch,  Karl von},
  year = {1993},
  month = dec 
}

@article{Haldane1954,
  title = {A statistical analysis of communication in “Apis mellifera” and a comparison with communication in other animals},
  volume = {1},
  ISSN = {1420-9098},
  url = {http://dx.doi.org/10.1007/BF02222949},
  DOI = {10.1007/bf02222949},
  number = {3},
  journal = {Insectes Sociaux},
  publisher = {Springer Science and Business Media LLC},
  author = {Haldane,  J. B. S. and Spurway,  H.},
  year = {1954},
  month = sep,
  pages = {247–283}
}

@article{Schurch2015,
  title = {The spatial information content of the honey bee waggle dance},
  volume = {3},
  ISSN = {1662-5161},
  url = {http://dx.doi.org/10.3389/fevo.2015.00022},
  DOI = {10.3389/fevo.2015.00022},
  journal = {Frontiers in Human Neuroscience},
  publisher = {Frontiers Media SA},
  author = {Sch\"urch,  Roger and Ratnieks,  Francis L. W.},
  year = {2015},
  month = mar 
}

@article{Sumpter2005,
  title = {The principles of collective animal behaviour},
  volume = {361},
  ISSN = {1471-2970},
  url = {http://dx.doi.org/10.1098/rstb.2005.1733},
  DOI = {10.1098/rstb.2005.1733},
  number = {1465},
  journal = {Philosophical Transactions of the Royal Society B: Biological Sciences},
  publisher = {The Royal Society},
  author = {Sumpter,  D.J.T},
  year = {2005},
  month = nov,
  pages = {5–22}
}

@article{Giardina2008,
  title = {Collective behavior in animal groups: Theoretical models and empirical studies},
  volume = {2},
  ISSN = {1955-2068},
  url = {http://dx.doi.org/10.2976/1.2961038},
  DOI = {10.2976/1.2961038},
  number = {4},
  journal = {HFSP Journal},
  publisher = {Informa UK Limited},
  author = {Giardina,  Irene},
  year = {2008},
  month = aug,
  pages = {205–219}
}

@article{Pratt2002,
  title = {Quorum sensing,  recruitment,  and collective decision-making during colony emigration by the ant Leptothorax albipennis},
  volume = {52},
  ISSN = {1432-0762},
  url = {http://dx.doi.org/10.1007/s00265-002-0487-x},
  DOI = {10.1007/s00265-002-0487-x},
  number = {2},
  journal = {Behavioral Ecology and Sociobiology},
  publisher = {Springer Science and Business Media LLC},
  author = {Pratt,  Stephen and Mallon,  Eamonn and Sumpter,  David and Franks,  Nigel},
  year = {2002},
  month = jul,
  pages = {117–127}
}

@article{Gordon2014,
  title = {The Ecology of Collective Behavior},
  volume = {12},
  ISSN = {1545-7885},
  url = {http://dx.doi.org/10.1371/journal.pbio.1001805},
  DOI = {10.1371/journal.pbio.1001805},
  number = {3},
  journal = {PLoS Biology},
  publisher = {Public Library of Science (PLoS)},
  author = {Gordon,  Deborah M.},
  year = {2014},
  month = mar,
  pages = {e1001805}
}

@article{Dall2005,
  title = {Information and its use by animals in evolutionary ecology},
  volume = {20},
  ISSN = {0169-5347},
  url = {http://dx.doi.org/10.1016/j.tree.2005.01.010},
  DOI = {10.1016/j.tree.2005.01.010},
  number = {4},
  journal = {Trends in Ecology \& Evolution},
  publisher = {Elsevier BV},
  author = {Dall,  S and Giraldeau,  L and Olsson,  O and McNamara,  J and Stephens,  D},
  year = {2005},
  month = apr,
  pages = {187–193}
}

@article{Sutton1998,
  title = {Reinforcement Learning: An Introduction},
  volume = {9},
  ISSN = {1941-0093},
  url = {http://dx.doi.org/10.1109/TNN.1998.712192},
  DOI = {10.1109/tnn.1998.712192},
  number = {5},
  journal = {IEEE Transactions on Neural Networks},
  publisher = {Institute of Electrical and Electronics Engineers (IEEE)},
  author = {Sutton,  R.S. and Barto,  A.G.},
  year = {1998},
  month = sep,
  pages = {1054–1054}
}

@article{Gold2007,
  title = {The Neural Basis of Decision Making},
  volume = {30},
  ISSN = {1545-4126},
  url = {http://dx.doi.org/10.1146/annurev.neuro.29.051605.113038},
  DOI = {10.1146/annurev.neuro.29.051605.113038},
  number = {1},
  journal = {Annual Review of Neuroscience},
  publisher = {Annual Reviews},
  author = {Gold,  Joshua I. and Shadlen,  Michael N.},
  year = {2007},
  month = jul,
  pages = {535–574}
}

@article{Wald1948,
  title = {Optimum Character of the Sequential Probability Ratio Test},
  volume = {19},
  ISSN = {0003-4851},
  url = {http://dx.doi.org/10.1214/aoms/1177730197},
  DOI = {10.1214/aoms/1177730197},
  number = {3},
  journal = {The Annals of Mathematical Statistics},
  publisher = {Institute of Mathematical Statistics},
  author = {Wald,  A. and Wolfowitz,  J.},
  year = {1948},
  month = sep,
  pages = {326–339}
}

@article{Bogacz2006,
  title = {The physics of optimal decision making: A formal analysis of models of performance in two-alternative forced-choice tasks.},
  volume = {113},
  ISSN = {0033-295X},
  url = {http://dx.doi.org/10.1037/0033-295x.113.4.700},
  DOI = {10.1037/0033-295x.113.4.700},
  number = {4},
  journal = {Psychological Review},
  publisher = {American Psychological Association (APA)},
  author = {Bogacz,  Rafal and Brown,  Eric and Moehlis,  Jeff and Holmes,  Philip and Cohen,  Jonathan D.},
  year = {2006},
  pages = {700–765}
}

@article{VelizCuba2016,
  title = {Stochastic Models of Evidence Accumulation in Changing Environments},
  volume = {58},
  ISSN = {1095-7200},
  url = {http://dx.doi.org/10.1137/15M1028443},
  DOI = {10.1137/15m1028443},
  number = {2},
  journal = {SIAM Review},
  publisher = {Society for Industrial \& Applied Mathematics (SIAM)},
  author = {Veliz-Cuba,  Alan and Kilpatrick,  Zachary P. and Josić,  Krešimir},
  year = {2016},
  month = jan,
  pages = {264–289}
}

@article{Dreller1998,
  title = {Division of labor between scouts and recruits: genetic influence and mechanisms},
  volume = {43},
  ISSN = {1432-0762},
  url = {http://dx.doi.org/10.1007/s002650050480},
  DOI = {10.1007/s002650050480},
  number = {3},
  journal = {Behavioral Ecology and Sociobiology},
  publisher = {Springer Science and Business Media LLC},
  author = {Dreller,  Claudia},
  year = {1998},
  month = aug,
  pages = {191–196}
}

@article{Liang2012,
  title = {Molecular Determinants of Scouting Behavior in Honey Bees},
  volume = {335},
  ISSN = {1095-9203},
  url = {http://dx.doi.org/10.1126/science.1213962},
  DOI = {10.1126/science.1213962},
  number = {6073},
  journal = {Science},
  publisher = {American Association for the Advancement of Science (AAAS)},
  author = {Liang,  Zhengzheng S. and Nguyen,  Trang and Mattila,  Heather R. and Rodriguez-Zas,  Sandra L. and Seeley,  Thomas D. and Robinson,  Gene E.},
  year = {2012},
  month = mar,
  pages = {1225–1228}
}

@book{Levathes2014,
  title={When China ruled the seas: The treasure fleet of the Dragon Throne, 1405--1433},
  author={Levathes, Louise},
  year={2014},
  publisher={Open Road Media}
}

@book{Bergreen2003,
  title={Over the Edge of the World: Magellan's Terrifying Circumnavigation of the Globe},
  author={Bergreen, Laurence},
  year={2003},
  publisher={William Morrow and Company}
}

@article{Charbonneau2015,
  title={Workers ‘specialized’ on inactivity: behavioral consistency of inactive workers and their role in task allocation},
  author={Charbonneau, Daniel and Dornhaus, Anna},
  journal={Behavioral ecology and sociobiology},
  volume={69},
  number={9},
  pages={1459--1472},
  year={2015},
  publisher={Springer}
}

@article{Conradt2012,
  title={Models in animal collective decision-making: information uncertainty and conflicting preferences},
  author={Conradt, Larissa},
  journal={Interface focus},
  volume={2},
  number={2},
  pages={226--240},
  year={2012},
  publisher={The Royal Society}
}

@article{Beshers2001,
  title={Models of division of labor in social insects},
  author={Beshers, Samuel N and Fewell, Jennifer H},
  journal={Annual review of entomology},
  volume={46},
  number={1},
  pages={413--440},
  year={2001},
  publisher={Annual Reviews 4139 El Camino Way, PO Box 10139, Palo Alto, CA 94303-0139, USA}
}

@article{Mann2018,
  title={Collective decision making by rational individuals},
  author={Mann, Richard P},
  journal={Proceedings of the National Academy of Sciences},
  volume={115},
  number={44},
  pages={E10387--E10396},
  year={2018},
  publisher={National Academy of Sciences}
}

@article{Mann2020,
  title={Collective decision-making by rational agents with differing preferences},
  author={Mann, Richard P},
  journal={Proceedings of the National Academy of Sciences},
  volume={117},
  number={19},
  pages={10388--10396},
  year={2020},
  publisher={National Academy of Sciences}
}

@article{Sumpter2008,
  title = {Information transfer in moving animal groups},
  volume = {127},
  ISSN = {1611-7530},
  url = {http://dx.doi.org/10.1007/s12064-008-0040-1},
  DOI = {10.1007/s12064-008-0040-1},
  number = {2},
  journal = {Theory in Biosciences},
  publisher = {Springer Science and Business Media LLC},
  author = {Sumpter,  David and Buhl,  Camille and Biro,  Dora and Couzin,  Iain},
  year = {2008},
  month = may,
  pages = {177–186}
}

@article{Biro2008,
  title={Are animal personality traits linked to life-history productivity?},
  author={Biro, Peter A and Stamps, Judy A},
  journal={Trends in ecology \& evolution},
  volume={23},
  number={7},
  pages={361--368},
  year={2008},
  publisher={Elsevier}
}

@article{Conradt2005,
  title={Consensus decision making in animals},
  author={Conradt, Larissa and Roper, Timothy J},
  journal={Trends in ecology \& evolution},
  volume={20},
  number={8},
  pages={449--456},
  year={2005},
  publisher={Elsevier}
}

@article{Jolles2017,
  title={Consistent individual differences drive collective behavior and group functioning of schooling fish},
  author={Jolles, Jolle W and Boogert, Neeltje J and Sridhar, Vivek H and Couzin, Iain D and Manica, Andrea},
  journal={Current Biology},
  volume={27},
  number={18},
  pages={2862--2868},
  year={2017},
  publisher={Elsevier}
}

@book{Topkis1998,
  title={Supermodularity and complementarity},
  author={Topkis, Donald M},
  year={1998},
  publisher={Princeton university press}
}

@article{Suganuma2025,
  title={How social learning enhances—or undermines—efficiency and flexibility in collective decision-making under uncertainty},
  author={Suganuma, Hidezo and Katahira, Kentaro and Ohtsuki, Hisashi and Kameda, Tatsuya},
  journal={Proceedings of the National Academy of Sciences},
  volume={122},
  number={48},
  pages={e2516827122},
  year={2025},
  publisher={National Academy of Sciences}
}

@article{Barendregt2025,
  title={Information-Seeking Decision Strategies Mitigate Risk in Dynamic, Uncertain Environments},
  author={Barendregt, Nicholas W and Gold, Joshua I and Josi{\'c}, Kre{\v{s}}imir and Kilpatrick, Zachary P},
  journal={arXiv preprint arXiv:2503.19107},
  year={2025}
}

@article{Bonabeau1996,
  title={Quantitative study of the fixed threshold model for the regulation of division of labour in insect societies},
  author={Bonabeau, Eric and Theraulaz, Guy and Deneubourg, Jean-Louis},
  journal={Proceedings of the Royal Society of London. Series B: Biological Sciences},
  volume={263},
  number={1376},
  pages={1565--1569},
  year={1996},
  publisher={The Royal Society London}
}

@article{Bonabeau1998,
  title = {Fixed Response Thresholds and the Regulation of Division of Labor in Insect Societies},
  volume = {60},
  ISSN = {0092-8240},
  url = {http://dx.doi.org/10.1006/bulm.1998.0041},
  DOI = {10.1006/bulm.1998.0041},
  number = {4},
  journal = {Bulletin of Mathematical Biology},
  publisher = {Springer Science and Business Media LLC},
  author = {Bonabeau,  E},
  year = {1998},
  month = jul,
  pages = {753–807}
}

@article{Theraulaz1998,
  title = {Response threshold reinforcements and division of labour in insect societies},
  volume = {265},
  ISSN = {1471-2954},
  url = {http://dx.doi.org/10.1098/rspb.1998.0299},
  DOI = {10.1098/rspb.1998.0299},
  number = {1393},
  journal = {Proceedings of the Royal Society of London. Series B: Biological Sciences},
  publisher = {The Royal Society},
  author = {Theraulaz,  G. and Bonabeau,  E. and Denuebourg,  J-N.},
  year = {1998},
  month = feb,
  pages = {327–332}
}

@article{Jeanson2014,
  title = {Interindividual variability in social insects – proximate causes and ultimate consequences},
  volume = {89},
  ISSN = {1469-185X},
  url = {http://dx.doi.org/10.1111/brv.12074},
  DOI = {10.1111/brv.12074},
  number = {3},
  journal = {Biological Reviews},
  publisher = {Wiley},
  author = {Jeanson,  Raphaël and Weidenm\"{u}ller,  Anja},
  year = {2014},
  month = dec,
  pages = {671–687}
}

@article{Ulrich2021,
  title={Response thresholds alone cannot explain empirical patterns of division of labor in social insects},
  author={Ulrich, Yuko and Kawakatsu, Mari and Tokita, Christopher K and Saragosti, Jonathan and Chandra, Vikram and Tarnita, Corina E and Kronauer, Daniel JC},
  journal={PLoS Biology},
  volume={19},
  number={6},
  pages={e3001269},
  year={2021},
  publisher={Public Library of Science San Francisco, CA USA}
}

\newpage

\setcounter{section}{0}
\renewcommand{\thesection}{SI}
\section{Supporting Information Appendix}
\subsection{Detailed model description}
\subsubsection*{Collective foraging in a changing environment} \label{sect:foraging}
We consider a group of $N$ agents with a single central home, which is situated at a distance from a single foraging site.
The placement of the home site is motivated by dwellings of socially foraging animal groups who bring food back to a central location (e.g., honey bees, ants, and many other social insects), serving as a base from which information and resources are collected. 
At each time $t = 1,2,\cdots$, each agent chooses to either idle or commit. Individuals who choose to idle remain at the home site, avoiding the expense and danger of exploration, but also foregoing potential rewards. In contrast, agents who choose to commit to the foraging site will leave the home site during a time step and may receive a reward but incur some cost for exploring, $\lambda$. Each agent can choose whether to commit or idle at each step. Rewards are resources that benefit the collective, such as food that is immediately consumed or nectar that is brought back to the home site. The cost of exploring includes both expended energy and  risks of foraging such as predation.

When visiting the site, agents receive a reward with a probability determined by the  hidden state of the environment at each time $t$. When  the environment is in the high (low) rewarding state  $s^t=s_+$ ($s^t=s_-$), each forager receives a fixed reward with probability $\gamma_+$ ($\gamma_-$), according to an independent and identically distributed binary random variable, conditioned on the state, 
\begin{equation*}
    \pr{x^t = 1 \middle| s^t = s_\pm, a^t = 1} = \gamma_\pm, \quad \gamma_- < \gamma_+, 
\end{equation*}
where $x^t = 1$ if a reward is received by a single agent ($a^t = 1$) and $x^t = 0$ otherwise. Agents that do not visit the site do not obtain rewards.

The environmental state changes probabilistically as a two-state Markov chain with symmetric switching (hazard) rates
\begin{equation*}
    \pr{s^{t+1} \neq s^t|s^t} = \epsilon,
\end{equation*}
where $\epsilon \in [0,1/2]$. Our results can be extended to the case in which transitions out of either state occur at different rates, $\pr{s^{t+1} = s_\pm | s^t = s_\mp} = \epsilon_{\pm}$, leading to unequal average dwell times (rather than the symmetric case considered here),
but this substantially complicates the resulting expressions.

\subsubsection*{Collective perception of environment} \label{sect:belief}

We assume agents combine their own observations with social information shared across the group to compute the posterior probability that the environment is in 
the high rewarding state~\cite{degroot2005optimal}. 
We also assume that all agents share their observations with all others once home, at which point everyone in the group has the same information and the same \emph{belief} about the environmental state. 
For example, in a sufficiently small colony of honey bees, all members can observe each other's waggle dances, which communicate information about foraging sites \cite{Frisch1993,Haldane1954,Schurch2015}. 
We derive a recursive expression for the probability that the environment is one of the two states given the observations of the collective, corresponding to the belief shared by all agents.

An agent infers the state, $s^t$, using the outcome of all foraging attempts made by other agents, and knowledge about the volatility of the environment, $\epsilon$. Let $a^t$ be the number of agents choosing to forage at time $t$. Among these agents, $x^t$ agents successfully receive rewards by finding food at the foraging site. The probability that $x^t$ out of $a^t$ agents receive rewards if the state is $s^t$ is given by
\begin{equation*}
    \pr{x^t | s^t = s_\pm, a^t} = \mathcal{B}(x^t | a^t, \gamma_\pm) \equiv {a^t \choose x^t}\gamma_\pm^{x^t} (1-\gamma_\pm)^{a^t - x^t}.
\end{equation*}
 We will discuss later how agents ideally choose to take specific actions based on their belief in Sect. \ref{sect:policy}, focusing on how the number $a^t$ of foragers on time step $t$ is chosen based on prior rewarded $x^{1:t-1} = \{x^1,\cdots,x^{t-1}\}$ and foraging $a^{1:t-1} = \{a^1,\cdots,a^{t-1}\}$ agent counts.
The best number $a^t$ of foragers to send out is in turn determined by the subjective probability (belief) that the current state is high-rewarding, $s^t = s_+$, based on these previous observations. We denote this belief by
\begin{equation*}
    g^t = \pr{s^t = s_+ | x^{1:t-1},a^{1:t-1}}.
\end{equation*}
Once $a^t$ agents commit, $x^t \in \{ 0, 1,..., a^t\}$ is sampled from a binomial distribution with parameters determined by the present state $s^t$. Subsequently all agents \emph{update} their belief based on $x^t$ and $a^t$. We can thus compute this update iteratively by accounting for this conditional dependence of $x^t$ and the remaining prior contributions to the belief:
\begin{align*}
    \pr{s^t | a^{1:t},x^{1:t}} &\propto \pr{x^t|s^t,a^t, a^{1:t-1},x^{1:t-1}}\pr{s^t,a^t |x^t,a^{1:t-1},x^{1:t-1}} \nonumber \\
    &\propto \pr{x^t| s^t,a^t} \pr{s^t | a^{1:t-1},x^{1:t-1}}. \nonumber
\end{align*}
Since there are only two potential values for the state, $s_\pm$, conservation of probability ensures
\begin{align*}
    \pr{s^t = s_+\middle|a^{1:t},x^{1:t}} &= \frac{\mathcal{B}(x^t|a^t,\gamma_+) g^t}{\mathcal{B}(x^t|a^t,\gamma_+) g^t + \mathcal{B}(x^t|a^t,\gamma_-) (1-g^t)} \nonumber \\
    &= \frac{g^t}{g^t + (1-g^t) \mathcal{C}(x^t,a^t)} \equiv \mathcal{U} (x^t|g^t,a^t),
\end{align*}
where $\mathcal{C}(x,a) = (\gamma_-/\gamma_+)^x ((1-\gamma_-)/(1-\gamma_+))^{a-x}$ is the ratio of the conditional binomial probabilities in either state. After this update, each agent \emph{predicts} the next state, $s^{t+1}$, from the obtained posterior probability (belief) of the current state. The Chapman-Kolmogorov equation for the state yields
\begin{align*}
    g^{t+1} = \pr{s^{t+1}|a^{1:t},x^{1:t}} &= \sum_{s^t} \pr{s^{t+1}|s^t,a^{1:t},x^{1:t}} \pr{s^t|a^{1:t},x^{1:t}} \nonumber \\
    &= \sum_{s^t} \pr{s^{t+1}|s^t} \pr{s^t|a^{1:t},x^{1:t}}, \nonumber
\end{align*}
according to the assumption that $s^{t+1}$ only depends on $s^t$. It follows that
\begin{equation} \label{eq:g^trans}
    g^{t+1} = (1-\epsilon) \mathcal{U}(x^t|g^t,a^t) + \epsilon(1-\mathcal{U}(x^t|g^t,a^t)) \equiv \mathcal{G}(x^t|g^t,a^t).
\end{equation}

We introduced a Bayesian model of collective perception in a changing environment. In this framework, both the agents' returns and their beliefs are shaped by their actions, $a^t$.
This raises a central question: given a belief, $g^t$, what is the optimal number of agents that should forage at each time step? In Sect. \ref{sect:policy}, we address this question by using dynamic programming to find the optimal policy of collective agents to maximize their return, and discuss how individuals without a coordinator optimize their decisions as a collective.

\subsubsection*{Normative foraging strategy} \label{sect:policy}

The allocation of agents impacts the immediate reward, and determines the information gathered (Section \ref{sect:belief}), and hence future rewards received by the group. The \emph{group action} at time $t$ is the number of foraging agents, denoted as $a^t$. Since the group size is constant, the number of idlers is $N - a^t$. Therefore, the group action (deploy $a^t$ foragers) is chosen from the action space:
\begin{equation*}
    a^t \in \mathcal{A} = \{0,1,\cdots,N\}.
\end{equation*}
Let $x^t$ represent the number of agents who receive a reward at time $t$ and let $g$ denote the probability that $s^t = s_+$ (belief). The expected immediate reward (per unit time), $\bar{r}^t$, for a given action is then expressed as:
\begin{align*}
    \bar{r}^t(g^t|a) &= \text{E}[r^t |a^t = a] \nonumber\\
    &= \text{E}[x^t - \lambda a^t | a^t = a] \nonumber \\
    &= a \left(\gamma_+ g^t + \gamma_- (1-g^t) - \lambda\right) \equiv a \bar{r}_1(g^t),
\end{align*}
where $\lambda$ is the commitment cost a single agent incurs during the trip, and $\bar{r}_1(g)$ is the expected reward for a single agent given belief $g$.
We assume that idling incurs no cost for simplicity, but could compute idling cost effects by shifting $\bar{r}^t$ and scaling $\lambda$ appropriately. 
From the viewpoint of a single agent, a \emph{greedy} strategy is one that always selects the action that gives the best immediate reward. 
Let $\theta_c$ be the \emph{critical threshold} where $\bar{r}_1(\theta_c) = 0$, which is given by
\begin{align*}
    \theta_c = \frac{\lambda - \gamma_-}{\gamma_+ - \gamma_-}.
\end{align*}
The greedy agent chooses to forage if $g > \theta_c$, and otherwise chooses to idle. 

In addition to potentially receiving a reward, the presence or absence of the reward also provides noisy evidence about the state of the environment.
The anticipated value of the information obtained from returning agents should also be taken into account when computing the value of an action. 
We do so by obtaining the probability of a future belief, $g^{t+1}$, resulting from 
sending out $a^t$ agents, given that the current belief is $g^t$. 
Given $g^t$ and $a^t$, there are at most $a^t +1$ possible values for $g^{t+1}$ since $x^t \in \{0,1,\cdots, a^t\}$. At time $t$, if $a^t = a$ agents have gone out to forage and $x^t = x$ of these agents receive rewards,~\eqref{eq:g^trans} gives the belief transition from $g^t = g$ to $g^{t+1} = g'$ as follows:
\begin{equation} \label{eq:belief^trans}
    p(g'|g,a) = \mathcal{B}(x|a,\gamma_+) g + \mathcal{B}(x|a,\gamma_-) (1-g), \quad g' = \mathcal{G}(x|g,a).
\end{equation}
Hence, the belief transition probability is determined completely by the probability of receiving $x$ rewards when sending out $a$ agents, given the current belief $g$.

We next determine the optimal foraging strategy, $\pi: [0,1] \to \mathcal{A}$ determining how many agents should be deployed given a belief, $g^t = g \in [0,1]$?
To measure foraging efficiency, we introduce the \emph{average reward}~\cite{Puterman1994,Mahadevan1996}.
Denote by $r^{t\pi}$ the collective reward at time $t$, given that the planner uses policy $\pi$.
Then the average reward in the long time limit is given by
\begin{equation*}
    \rho^\pi = \lim_{T \to \infty} \frac{1}{T}\text{E}\left[\sum_{t=1}^T r^{t \pi}\right],
\end{equation*}
which is independent of the initial action if the underlying Markov decision process is ergodic \cite{Mahadevan1996}. This average reward satisfies  the Bellman equation,
\begin{equation} \label{eq:bellman}
    V^\pi(g) + \rho^\pi = r(g|\pi(g)) + \mathcal{P}V^\pi(g|\pi(g)), \quad \mathcal{P}V(g|a) = \sum_{g'} V(g') p(g'|g,a),
\end{equation}
which incorporates the adjusted value function
\begin{equation*}
    V^\pi(g) = \text{E} \left[\sum_{s = t}^\infty r^{s \pi} - \rho^\pi \middle| g^t = g\right].
\end{equation*}
On the right-hand side of \eqref{eq:bellman}, the first term represents the expected immediate reward and the second term accounts for the expected future rewards if $a = \pi(g)$ agents are sent out to forage under policy $\pi$.  The probability $p(g'|g,a)$ is defined by~\eqref{eq:belief^trans}, and the sum is over the $N+1$
possible outcomes of the belief, $g'$, at the next timestep.
Using dynamic programming, we can determine the optimal number of agents that should commit to the environment given the current belief $g$, denoted by the policy $\pi(g)$, which satisfies Bellman's optimality equation:
\begin{equation} \label{eq:bellman_opt}
    V(g) + \rho = \max_{a = 0,1,\cdots,N} \left\{\ r(g|a) + \mathcal{P}V(g|a) \right\}.
\end{equation}

\subsection{Asymptotic analysis of Bellman equation}
\subsubsection*{Two-step look-ahead approximation}

We now consider the two-step look-ahead model, whose value function is given by
\begin{equation} \label{eq:two-step}
    V_\text{two-step}(g|a) = a r(g) + \mathcal{P} V_\text{one-step}(g|a), \quad V_\text{one-step}(g) = N r(g) H(g-\theta_c),
\end{equation}
where $H(g)$ denotes the Heaviside function, equal to $1$ if $g \ge 0$, and $0$ otherwise.
Here the first term represents the immediate reward, and the second term represents the expected return by looking two-step ahead. The one-step-ahead value function is determined by the greedy strategy. The optimal composition in the two-step-ahead model satisfies
\begin{equation*}
    \pi_\text{two-step}(g) = \underset{a = 0,1,\cdots,N}{\arg\max} V_\text{two-step}(g|a).
\end{equation*}

We next determine the first two leading-order terms of the second term in the value function, $\mathcal{P} V_\text{one-step}(g|a)$ by expanding as
\begin{align*}
    \frac{1}{N}\mathcal{P}V_\text{one-step}(g|a) &= \sum_{\mathcal{G}(x|g,a) \geq \theta_c} (\mathcal{G}(x|g,a) - \theta_c) p(x|g,a) \nonumber \\
    &=(1-\epsilon - \theta_c) g \sum_{\mathcal{G}(x|g,a) \geq \theta_c} \mathcal{B}(x|a,\gamma_+) - (\theta_c - \epsilon)(1-g) \sum_{\mathcal{G}(x|g,a) \geq \theta_c} \mathcal{B}(x|a,\gamma_-) \nonumber \\
    &:= (1-\epsilon - \theta_c) g \big(1 - \mathcal{R}_+(a|g)\big) - (\theta_c - \epsilon)(1-g) \mathcal{R}_-(a|g),
\end{align*}
based on the identity
\begin{equation*}
    \mathcal{G}(x|g,a)p(x|g,a) = (1-\epsilon) g \mathcal{B}(x|a,\gamma_+) + \epsilon (1-g) \mathcal{B}(x|a,\gamma_-).
\end{equation*}
We analyze the remainders $\mathcal{R}_\pm$ using large deviation theory \cite{Dembo2010,Bressloff2021}.
For simplicity, we again assume $\gamma_+ + \gamma_- = 1$ and denote $\gamma_+ = \gamma$. Let $X_{a,\gamma}$ denote a binomial random variable with size $a$ and probability $\gamma$. We focus on the non-trivial regime $\epsilon < \theta_c <1-\epsilon$, so that
\begin{align*}
    \mathcal{R}_+ (a|g) &= \text{Pr}\big( \mathcal{G}(X_{a,\gamma_+}|g,a) < \theta_c\big) \nonumber\\
    &= \text{Pr}\left(X_{a,\gamma_+} < \frac{a}{2} + z(g)\right), \quad z(g) = \frac{\log \frac{\theta_c - \epsilon}{1-\epsilon - \theta_c} - \log \frac{g}{1-g}}{\log \frac{\gamma}{1-\gamma}}, \nonumber\\
    &= \text{Pr} \left(X_{a,\gamma_-} > \frac{a}{2} - z(g)\right).
\end{align*}
Similarly, 
\begin{align*}
    \mathcal{R}_-(a|g) = \text{Pr}\left(X_{a,\gamma_-} \geq \frac{a}{2} + z(g)\right).
\end{align*}
By Cramer's theorem, the binomial random variable satisfies the large deviation principle:
\begin{align} \label{eq:cramer}
    \lim_{a \to \infty} \frac{1}{a} \ln \text{Pr} (X_{a,\gamma} > \xi a) = - \mathcal{I}(\xi|\gamma), \quad \mathcal{I}(\xi|\gamma) = \xi \ln \frac{\xi}{\gamma} + (1-\xi) \ln \frac{1-\xi}{1-\gamma},
\end{align}
for $\xi > \gamma$, where $\mathcal{I} >0 $ is the rate function for the binomial random variable. For sufficiently large $a$ and any $0 < \zeta < \frac{1}{2} - (1-\gamma) = \gamma - 1/2$, the following inequality holds
\begin{align*}
    \left(\frac{1}{2} - \zeta\right)a <\frac{a}{2} - |z(g)| \leq \frac{a}{2} + |z(g)| < \left(\frac{1}{2} + \zeta\right)a
\end{align*}
which implies that
\begin{align*}
    \text{Pr}\left(X_{a,1-\gamma} > \left(\frac{1}{2} + \zeta\right) a \right) \leq \mathcal{R}_\pm (a|g) \leq \text{Pr}\left(X_{a,1-\gamma} > \left(\frac{1}{2} - \zeta\right) a \right).
\end{align*}
Applying the binomial rate function gives
\begin{align*}
    - \mathcal{I}\left(\frac{1}{2} - \zeta\middle|1-\gamma\right) \leq \lim_{a \to \infty}\frac{1}{a}\ln \mathcal{R}_\pm(a|g) \leq - \mathcal{I}\left(\frac{1}{2} + \zeta\middle|1-\gamma\right).
\end{align*}
Since $\zeta > 0$ is arbitrary, it follows that
\begin{align*}
    \lim_{a \to \infty} \frac{1}{a} \ln \mathcal{R}_\pm (a|g) = - \mathcal{I}_0 := - \mathcal{I}\left(\frac{1}{2} \middle| 1 -\gamma\right).
\end{align*}
Thus, the operator $\mathcal{P}$ admits the asymptotic form:
\begin{equation} \label{eq:two-step_expansion}
    \frac{1}{N}\mathcal{P}V_\text{one-step}(g|a) \sim \mathcal{B}_0(g) - c\mathcal{B}_1(g) e^{-a\mathcal{I}_0},
\end{equation}
for sufficiently large $a$, where $c$ is a positive constant and 
\begin{equation*}
    \mathcal{B}_0(g) = g ( 1- \epsilon - \theta_c ) - (1-g) (\theta_c - \epsilon), \quad \mathcal{B}_1(g) = g ( 1-\epsilon - \theta_c) + (1-g) (\theta_c - \epsilon).
\end{equation*}
Note that $\mathcal{B}_1 > 0$ for any $g \in (\epsilon,1-\epsilon)$.

We finally optimize the value function in~\eqref{eq:two-step} by substituting the expansion from \eqref{eq:two-step_expansion}:
\begin{align*}
    V_\text{two-step}(g|a) = a r(g) + N \big( \mathcal{B}_0(g) - c \mathcal{B}_1(g) e^{- a \mathcal{I}_0}\big).
\end{align*}
If $g \geq \theta_c$, the value function $V_\text{two-step}(g|a)$ increases with $a$, and is therefore maximized at $a = N$. Otherwise, if $g < \theta_c$, there exists a value $a < N$ satisfying the critical condition
\begin{align*}
    r(g) + c N \mathcal{I}_0\mathcal{B}_1(g) e^{-a\mathcal{I}_0} = 0.
\end{align*}
Solving for $a$ gives
\begin{align*}
    a = \frac{\log N}{\mathcal{I}_0} + \log \mathcal{I}_0 + \log \frac{c \mathcal{A}_1(g)}{\theta_c - g} + \cdots,
\end{align*}
Therefore, the optimal composition in the two-step-ahead model is
\begin{align*}
    \pi(g) \sim \begin{cases}
        \mathcal{O}(\log N) &,~ g < \theta_c \\
        N &,~ g \geq \theta_c
    \end{cases}.
\end{align*}
for sufficiently large $N$. Put simply, nearly all agents should refrain from foraging unless the belief $g \geq \theta_c$ and a small, logarithmically scaled minority should forage at lower beliefs.

\subsubsection*{First-order optimal policy}
The optimal composition of individual strategies satisfies  Bellman's equation given in~\eqref{eq:bellman}:
\begin{equation*}
    V(g) + \rho = \max_{a = 0,1,\cdots,N} \left\{ ar(g) + \mathcal{P}V(g|a)\right\},
\end{equation*}
where $\mathcal{P}$ denotes the operator for the expected future value, 
\begin{equation*}
    \mathcal{P}V(g|a) = \sum_{x = 0}^a (V\circ \mathcal{G})(x|g,a) p(x|g,a).
\end{equation*}
Here $\mathcal{G}(x|g,a)$ denotes the updated belief given current belief $g$, the number of committed foragers $a$, and the number of successful foragers $x$. The term $p(x|g,a)$ represents the corresponding probability. For simplicity, we assume $\gamma_+ + \gamma_- = 1$ and denote $\gamma_+ = \gamma$. Taking $a \to \infty$, we have
\begin{align}
    \lim_{a \to \infty} \mathcal{G}(x|g,a) &= \lim_{a \to \infty}\frac{(1-\epsilon) g + \epsilon (1-g) C_{x,a}}{g + (1-g) C_{x,a}}, \quad C_{x,a} = \left(\frac{\gamma}{1-\gamma}\right)^{a - 2x}, \nonumber \\
    & \to \begin{cases}
        \epsilon &,~ x < a/2 \\
        (1-\epsilon) g + \epsilon (1-g) &,~ x = a/2 \\
        1- \epsilon &,~ x > a/2
    \end{cases}. \label{eq:Gconv}
\end{align}
We can also compute the probability limit
\begin{align*}
    \lim_{a \to \infty} \sum_{x < a/2} p(x|g,a) &= \lim_{a \to \infty} g \sum_{x < a/2} \mathcal{B}(x|a,\gamma_+) + (1-g) \sum_{x<a/2}\mathcal{B}(x|a,\gamma_-) = 1-g.
\end{align*}
Similarly,
\begin{align*}
    \lim_{a \to \infty} \sum_{x > a/2} p(x|g,a) = g.
\end{align*}
Using these limits, the operator converges to
\begin{align*}
    \lim_{a \to \infty} \mathcal{P}V(g|a) &= \lim_{a \to \infty} V(\epsilon) \sum_{x<a/2} p(x|g,a) + V(1-\epsilon) \sum_{x > a/2} p(x|g,a) \nonumber \\
    &= g V(1-\epsilon) + (1-g) V(\epsilon) := \mathcal{P}_0 V(g). 
\end{align*}
Introducing the expansion
\begin{equation*}
    a = \pi(g) = N\pi_0(g) + \cdots, \quad \pi_0(g) \in [0,1],
\end{equation*}
we find that the leading-order behavior of the operator is given by
\begin{align*}
    \mathcal{P}V(g|a) = \mathcal{P}_0V(g) + \cdots.
\end{align*}
We further introduce the expansions
\begin{equation*}
    V(g|a) = NV_0(g|a) + \cdots, \ \ \ \ \ \rho = N \rho_0 + \cdots.
\end{equation*}
Substituting those expansions into Bellman's equation yields
\begin{align*}
    N V_0(g|a) + N \rho_0 + \cdots = \max \left\{N\pi_0(g) r(g) + \mathcal{P}_0 V_0(g) + \cdots\right\}.
\end{align*}
Collecting $\mathcal{O}(N)$ terms gives the leading-order equation
\begin{align*}
    V_0(g|a) + \rho_0 &= \max \{ \pi_0(g) r(g) + \mathcal{P}_0 V_0(g) \} \nonumber \\
    &= \max \{ \pi_0(g) r(g)\} + \mathcal{P}_0V_0(g),
\end{align*}
since $\mathcal{P}_0$ is independent of $a$. Since $r(g) \geq 0$ when $g \geq \theta_c$, the right-hand side is maximized at $\pi_0(g) = 1$. Similarly, it is maximized at $\pi_0(g) = 0$ if $g< \theta_c$. Thus, the leading-order optimal proportion of committed foragers is given by
\begin{align*}
    \pi_0(g) = H(g-\theta_c),
\end{align*}
where $H(g)$ is the Heaviside function, equal to $1$ if $g\geq 0$ and $0$ otherwise. It follows that
\begin{align*}
    \pi(g) = N H(g-\theta_c) + \cdots,
\end{align*}
which yields $\pi(g) \sim N H(g - \theta_c)$, therefore, all individuals should commit when $g \geq \theta_c$.

\subsubsection*{Upper bound on the optimal policy when $g < \theta_c$}

We now establish that the optimal policy $\pi(g) \leq \Pi_N(g) =  \mathcal{O}(\ln N)$ when $g < \theta_c$. The proof proceeds in four steps: (i) establishing pointwise convergence of the posterior update, (ii) bounding the difference between the finite and limiting operators, (iii) showing the relevant constants scale linearly in $N$, and (iv) deriving the upper bound on the optimal action.

\paragraph{Pointwise convergence of the posterior update.}
Define the rescaled posterior update $\tilde{\mathcal{G}}(\xi|g,a) = \mathcal{G}(\xi a|g,a)$ for $\xi \in [0,1]$. The pointwise limit as $a \to \infty$ is given by
\begin{equation}\label{eq:Glimit}
    \tilde{\mathcal{G}}_\infty(\xi|g) := \lim_{a \to \infty}\tilde{\mathcal{G}}(\xi|g,a) = \begin{cases}
        \epsilon &,~ \xi < 1/2, \\
        (1-\epsilon) g + \epsilon (1-g) &,~ \xi = 1/2, \\
        1- \epsilon &,~ \xi > 1/2.
    \end{cases}
\end{equation}
We now quantify the rate of convergence. For $\xi < 1/2$, we have
\begin{align*}
     \tilde{\mathcal{G}}(\xi|g,a) - \tilde{\mathcal{G}}_\infty (\xi|g) = (1- 2\epsilon) \tilde{\mathcal{U}}(\xi|g,a),
\end{align*}
where $\tilde{\mathcal{U}}(\xi|g,a) = \mathcal{U}(\xi a|g,a)$ satisfies
\begin{align*}
    \tilde{\mathcal{U}}(\xi|g,a) = \frac{g\mathcal{Q}_\xi^{-a}}{g \mathcal{Q}_\xi^{-a} + (1-g)}, \qquad \mathcal{Q}_\xi := \left(\frac{\gamma}{1-\gamma}\right)^{|2\xi -1|}.
\end{align*}
Here we assume $\gamma > 1/2$, so that $\mathcal{Q}_\xi > 1$ for $\xi \neq 1/2$.
Since $g \leq \theta_c < 1$, we obtain the bound
\begin{align*}
    \tilde{\mathcal{U}}(\xi|g,a) &\leq \frac{g}{1-g} \mathcal{Q}_\xi^{-a} \leq \frac{\theta_c}{1- \theta_c} \exp\left(-a\ln\mathcal{Q}_\xi\right).
\end{align*}
Therefore,
\begin{equation}\label{eq:Gbound}
    \left|\tilde{\mathcal{G}}(\xi|g,a) - \tilde{\mathcal{G}}_\infty (\xi|g)\right| \leq Z \exp\left(-a\ln\mathcal{Q}_\xi\right), \qquad Z := \frac{(1-2\epsilon) \theta_c}{1-\theta_c}.
\end{equation}
A symmetric argument establishes the same bound for $\xi > 1/2$. Note that the decay rate $\ln \mathcal{Q}_\xi \to 0$ as $\xi \to 1/2$, reflecting slower convergence near $x = a/2$.

\paragraph{Asymptotic analysis of the operator $\mathcal{P}$.}
Define the limiting operator $\mathcal{P}_\infty V(g) := \lim_{a \to \infty} \mathcal{P}V(g|a)$ and the difference
\[
\Delta \mathcal{P}V(g|a) := \mathcal{P}V(g|a) - \mathcal{P}_\infty V(g).
\]
Fix a parameter $\zeta \in (0,\gamma -  1/2)$ and introduce the set
\begin{align*}
    \Lambda(a,\zeta) := \left\{ x \in \{0,1,\ldots,a\} ~\middle|~ \left|\frac{x}{a} - \frac{1}{2}\right| \leq \zeta\right\},
\end{align*}
which captures observations near the boundary $x/a = 1/2$.
We decompose the difference as
\begin{align*}
    \Delta \mathcal{P}V(g|a) &= \sum_{x = 0}^a \left[(V\circ\tilde{\mathcal{G}})\left(\frac{x}{a}\Big|g,a\right) - (V\circ\tilde{\mathcal{G}}_\infty)\left(\frac{x}{a}\Big|g\right)\right]p(x|g,a) \nonumber \\
    &= \Sigma_1 + \Sigma_2,
\end{align*}
where $\Sigma_1$ sums over $x \in \Lambda(a,\zeta)$ and $\Sigma_2$ sums over $x \notin \Lambda(a,\zeta)$.

The distribution $p(x|g,a)$ is a mixture of two binomials:
\begin{align*}
    p(x|g,a) = g \, B(x|a,\gamma) + (1-g) \, B(x|a,1-\gamma),
\end{align*}
where $B(x|a,p) = \binom{a}{x} p^x (1-p)^{a-x}$.
Since $\gamma > 1/2$, the first component $B(x|a,\gamma)$ has mean $\gamma a > a/2$, while the second component $B(x|a,1-\gamma)$ has mean $(1-\gamma)a < a/2$. Both components concentrate away from $x = a/2$.
By Cram\'{e}r's theorem applied to each component separately:
\begin{align*}
    \sum_{x \in \Lambda(a,\zeta)} B(x|a,\gamma) &\leq B_1 \exp\left(-a \, \mathcal{I}\left(\tfrac{1}{2}-\zeta \,\big|\, 1-\gamma\right)\right), \\
    \sum_{x \in \Lambda(a,\zeta)} B(x|a,1-\gamma) &\leq B_2 \exp\left(-a \, \mathcal{I}\left(\tfrac{1}{2}+\zeta \,\big|\, \gamma\right)\right),
\end{align*}
where $\mathcal{I}(y|p) = y \ln\frac{y}{p} + (1-y)\ln\frac{1-y}{1-p}$ is the Kullback-Leibler divergence (rate function for binomial large deviations), and $B_1, B_2$ are constants.
Define
\begin{align*}
    I_0 := \min\left\{ \mathcal{I}\left(\tfrac{1}{2}-\zeta \,\big|1-\, \gamma\right), \, \mathcal{I}\left(\tfrac{1}{2}+\zeta \,\big|\,\gamma\right) \right\} > 0.
\end{align*}
Note that $I_0 > 0$ according to \eqref{eq:cramer}. 
Therefore,
\begin{align*}
    \sum_{x \in \Lambda(a,\zeta)} p(x|g,a) &= g \sum_{x \in \Lambda(a,\zeta)} B(x|a,\gamma) + (1-g) \sum_{x \in \Lambda(a,\zeta)} B(x|a,1-\gamma) \nonumber \\
    &\leq g B_1 \exp(-a I_0) + (1-g) B_2 \exp(-a I_0) \nonumber \\
    &\leq B \exp(-a I_0),
\end{align*}
where $B = \max\{B_1, B_2\}$.
Thus,
\begin{equation}\label{eq:Sigma1}
    |\Sigma_1| \leq 2\|V\|_\infty \sum_{x \in \Lambda(a,\zeta)}p(x|g,a) \leq 2 B\|V\|_\infty  \exp \left(-a I_0\right).
\end{equation}

For $x \notin \Lambda(a,\zeta)$, we have $|x/a - 1/2| > \zeta$, so the bound \eqref{eq:Gbound} applies with $\ln \mathcal{Q}_{x/a} \geq \ln \mathcal{Q}_\zeta > 0$, where
\[
\mathcal{Q}_\zeta = \left(\frac{\gamma}{1-\gamma}\right)^{2\zeta}.
\]
Assuming $V$ is Lipschitz continuous with constant $\|V'\|_\infty$, we have
\begin{align*}
    \left|(V\circ\tilde{\mathcal{G}})\left(\frac{x}{a}\Big|g,a\right) - (V\circ\tilde{\mathcal{G}}_\infty)\left(\frac{x}{a}\Big|g\right)\right| 
    &\leq \|V'\|_\infty \left|\tilde{\mathcal{G}}\left(\frac{x}{a}\Big|g,a\right) - \tilde{\mathcal{G}}_\infty\left(\frac{x}{a}\Big|g\right)\right| \nonumber\\
    &\leq \|V'\|_\infty \cdot Z \exp\left(-a \ln \mathcal{Q}_\zeta\right).
\end{align*}
Summing over $x \notin \Lambda(a,\zeta)$ and using $\sum_x p(x|g,a) = 1$,
\begin{equation}\label{eq:Sigma2}
    |\Sigma_2| \leq \|V'\|_\infty \cdot Z \exp\left(-a \ln \mathcal{Q}_\zeta\right).
\end{equation}

Combining \eqref{eq:Sigma1} and \eqref{eq:Sigma2}, we obtain
\begin{equation}\label{eq:DeltaP}
    |\Delta \mathcal{P}V(g|a)| \leq D(V) \exp\left(-a \tilde{I}\right),
\end{equation}
where
\[
\tilde{I} := \min\{I_0, \ln \mathcal{Q}_\zeta\} > 0, \qquad D(V) := 2B\|V\|_\infty + Z\|V'\|_\infty.
\]

\paragraph{Scaling of $D(V)$ with $N$.}
Consider the ergodic Bellman equation
\begin{equation}\label{eq:bellman_scaled}
    V(g) + \rho = \max_{a \in \{0,1,\ldots,N\}} \left\{ -a\delta(g) + \mathcal{P}V(g|a)\right\},
\end{equation}
where $\delta(g) := -r(g) > 0$ for $g < \theta_c$.
Introduce the scaled quantities
\[
U(g) := \frac{V(g)}{N}, \qquad \rho' := \frac{\rho}{N}.
\]
Then $U$ satisfies
\begin{align*}
    U(g) + \rho' = \max_{a \in \{0,1,\ldots,N\}} \left\{ -\frac{a\delta(g)}{N} + \mathcal{P}U(g|a)\right\}.
\end{align*}
The immediate reward is uniformly bounded:
\begin{align*}
    \left|\frac{a\delta(g)}{N}\right| \leq \sup_{g} |\delta(g)| =: C_0 < \infty.
\end{align*}
By standard results on the span of the bias function in ergodic MDPs (see, e.g., \cite{Puterman1994}), we have
\begin{align*}
    \mathrm{span}(U) := \max_g U(g) - \min_g U(g) \leq \frac{2C_0}{1 - \beta},
\end{align*}
where $\beta < 1$ is a contraction coefficient depending on the mixing properties of the chain. Normalizing $U$ so that $\min_g U(g) = 0$, we obtain $\|U\|_\infty \leq 2C_0/(1-\beta)$.
For the derivative bound, differentiating the Bellman equation and using regularity of $\mathcal{P}$ with respect to $g$, one can show (under appropriate smoothness assumptions on the model) that
\begin{align*}
    \|U'\|_\infty \leq C_1,
\end{align*}
for some constant $C_1$ independent of $N$.
Rescaling back to $V$:
\begin{align*}
    \|V\|_\infty \leq C_2 N, \qquad \|V'\|_\infty \leq C_1 N,
\end{align*}
where $C_2 = 2C_0/(1-\beta)$. Therefore,
\begin{equation}\label{eq:Dbound}
    D(V) = 2B\|V\|_\infty + Z\|V'\|_\infty \leq (2BC_2 + ZC_1)N =: \tilde{Z} N.
\end{equation}

\paragraph{Upper bound on the optimal policy.}
Define the Bellman operator for a fixed action:
\begin{align*}
    \mathcal{T}V(g|a) := -a\delta(g) + \mathcal{P}V(g|a).
\end{align*}
The value difference from increasing the commitment level is
\begin{align*}
    \mathcal{T}V(g|a+1) - \mathcal{T}V(g|a) &= -\delta(g) + \mathcal{P}V(g|a+1) - \mathcal{P}V(g|a) \nonumber \\
    &= -\delta(g) + \left(\Delta \mathcal{P}V(g|a+1) - \Delta \mathcal{P}V(g|a)\right),
\end{align*}
where we used $\mathcal{P}_\infty V(g|a+1) = \mathcal{P}_\infty V(g|a) = \mathcal{P}_\infty V(g)$.
By the triangle inequality and \eqref{eq:DeltaP}--\eqref{eq:Dbound},
\begin{align*}
    \left|\Delta \mathcal{P}V(g|a+1) - \Delta \mathcal{P}V(g|a)\right| \leq 2D(V) \exp(-a\tilde{I}) \leq 2\tilde{Z} N \exp(-a\tilde{I}).
\end{align*}
Define the threshold
\begin{align*}
    \Pi_N(g) := \frac{1}{\tilde{I}}\left(\ln N + \ln\frac{4\tilde{Z}}{\delta(g)}\right).
\end{align*}
For any $a > \Pi_N(g)$, we have
\[
2\tilde{Z} N \exp(-a\tilde{I}) < \frac{\delta(g)}{2},
\]
and therefore
\begin{align*}
    \mathcal{T}V(g|a+1) - \mathcal{T}V(g|a) \leq -\delta(g) + \frac{\delta(g)}{2} = -\frac{\delta(g)}{2} < 0.
\end{align*}
This shows that increasing $a$ beyond $\Pi_N(g)$ strictly decreases the value, so the optimal policy satisfies
\begin{align*}
    \pi^*(g) \leq \Pi_N(g) = \frac{\ln N}{\tilde{I}} + \mathcal{O}(1).
\end{align*}
We conclude that $\pi^*(g) \leq \mathcal{O}(\ln N)$ for all $g < \theta_c$.

\subsection{Probabilistic decision-making in homogeneous collectives}
Consider a collective of $N$ individuals who independently follow the same probabilistic strategy $\varphi(g)$, defined as the probability of committing given belief $g$. The corresponding Bellman equation is given by
\begin{align}
    V(g) + \rho = \sum_{a = 0}^N \mathcal{B}\big(a\big|N,\varphi(g)\big) \big(r(g|a )+ \mathcal{P}V(g|a)\big),
\end{align}
where $a$ is the number of committed individuals and $\mathcal{B}(a|n,p)$ is the Binomial distribution with $n$ trials and success probability $p$. As before, $r(g|a)$ is the expected reward and $\mathcal{P}V(g|a)$ the future expected return given belief $g$ and action (or committed individuals) $a$. 
Note that the Bellman optimality equation can be written as
\begin{align}
    V(g) + \rho = \sum_{a = 0}^N 1_{a = \pi(g)} \big(r(g|a )+ \mathcal{P}V(g|a)\big),
\end{align}
where $1_{a = \pi(g)}$ represents the indicator function, equal to $1$ if and only if $a = \pi(g)$.
Even when $\varphi(g) = \pi(g)$, randomness in individual decisions lead to random over- or under-commitment relative to the optimal decision at the group level.
However, for the homogeneous collective, the induced binomial distribution satisfies
\begin{equation}
    \mathcal{B}\big(a\big|N,\pi(g)\big) \to^p 1_{a = \pi(g)}
\end{equation}
as $N \to \infty$, by the law of large numbers. In other words, despite individual-level randomness, the collective behavior concentrates on the optimal group action and becomes asymptotically near-optimal as group size $N$ increases.







\newpage
\setcounter{figure}{0}
\renewcommand{\thefigure}{SI.\arabic{figure}}
\begin{figure}
    \centering
    \includegraphics[width=\linewidth]{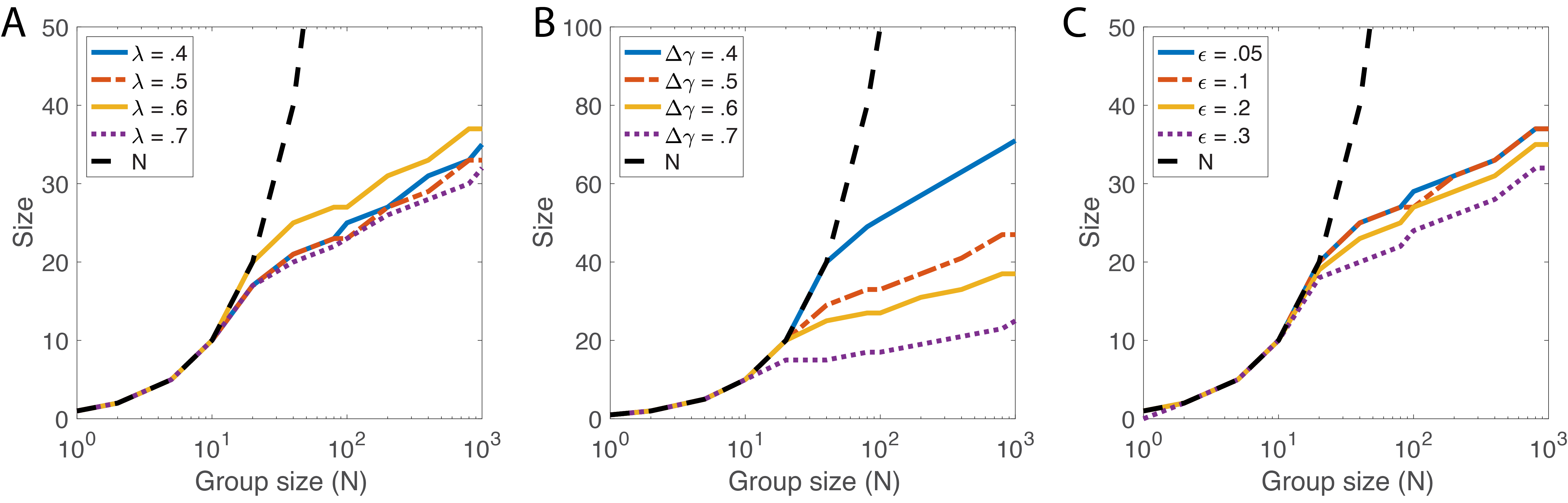}
    \caption{\textbf{Sublinear scaling of risk-tolerant individuals across ecological conditions.} 
    \textbf{(A)}~The number of risk-tolerant individuals as a function of group size $N$ for different commitment costs, $\lambda$.
    \textbf{(B)}~Scaling of risk-tolerant individuals with group size under varying reward conditions, $\Delta \gamma = \gamma_+ - \gamma_-$ (difference in high- and low-reward probabilities).
    \textbf{(C)}~Scaling under different environmental switching rates, $\epsilon$.
    Across all conditions, most individuals are risk-tolerant in small groups, whereas in larger groups the number of risk-takers increases sublinearly and logarithmically with group size.}
\end{figure}

\begin{figure}
    \centering
    \includegraphics[width=.4\linewidth]{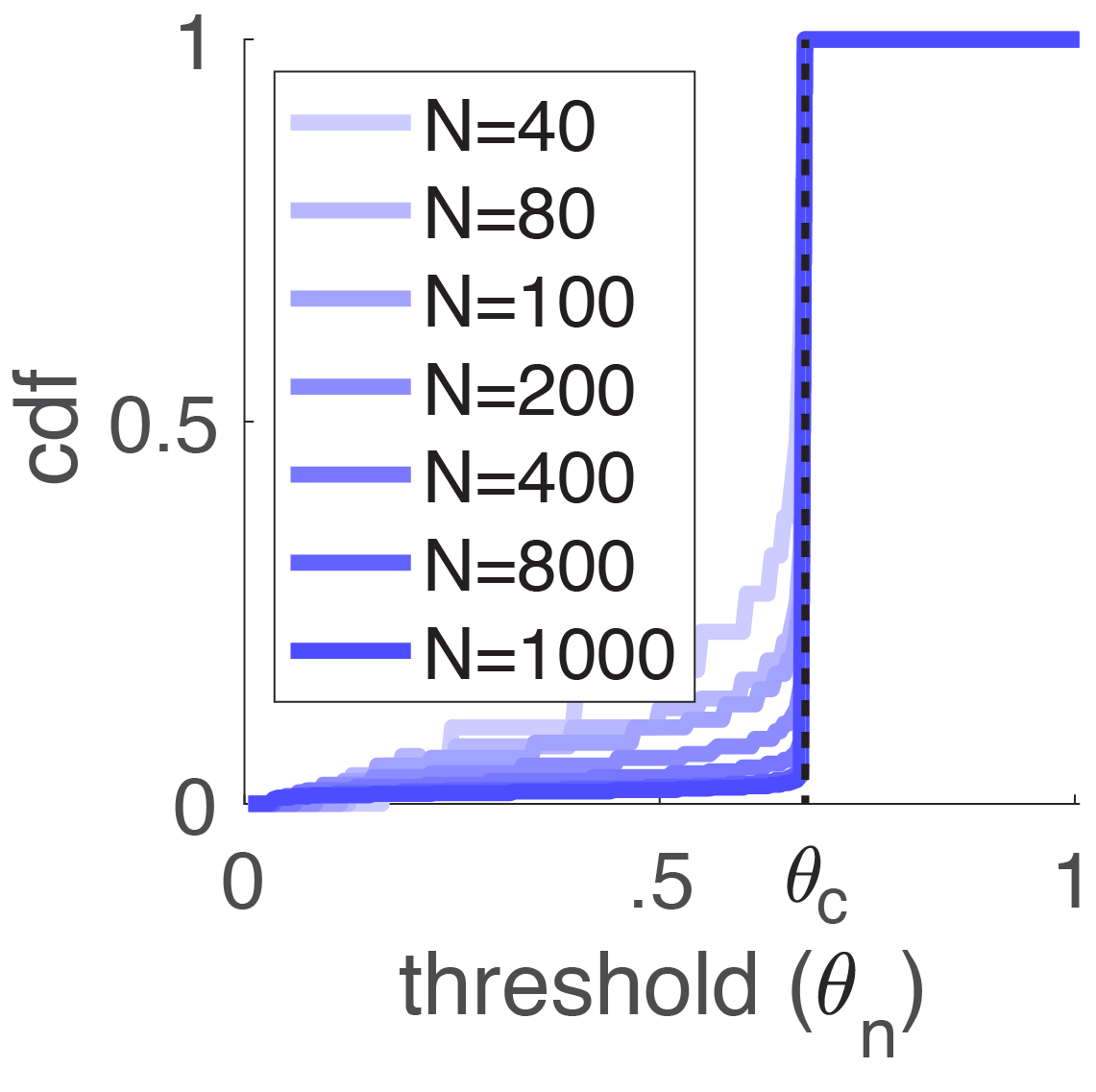}
    \caption{
    \textbf{Large collectives fully commit once expected rewards become positive.}
    The cumulative distribution of optimal thresholds indicates that all individuals commit when $g \geq \theta_c$, i.e., when the expected reward is positive, in sufficiently large collectives.}
\end{figure}

\begin{figure}
    \centering
    \includegraphics[width=\linewidth]{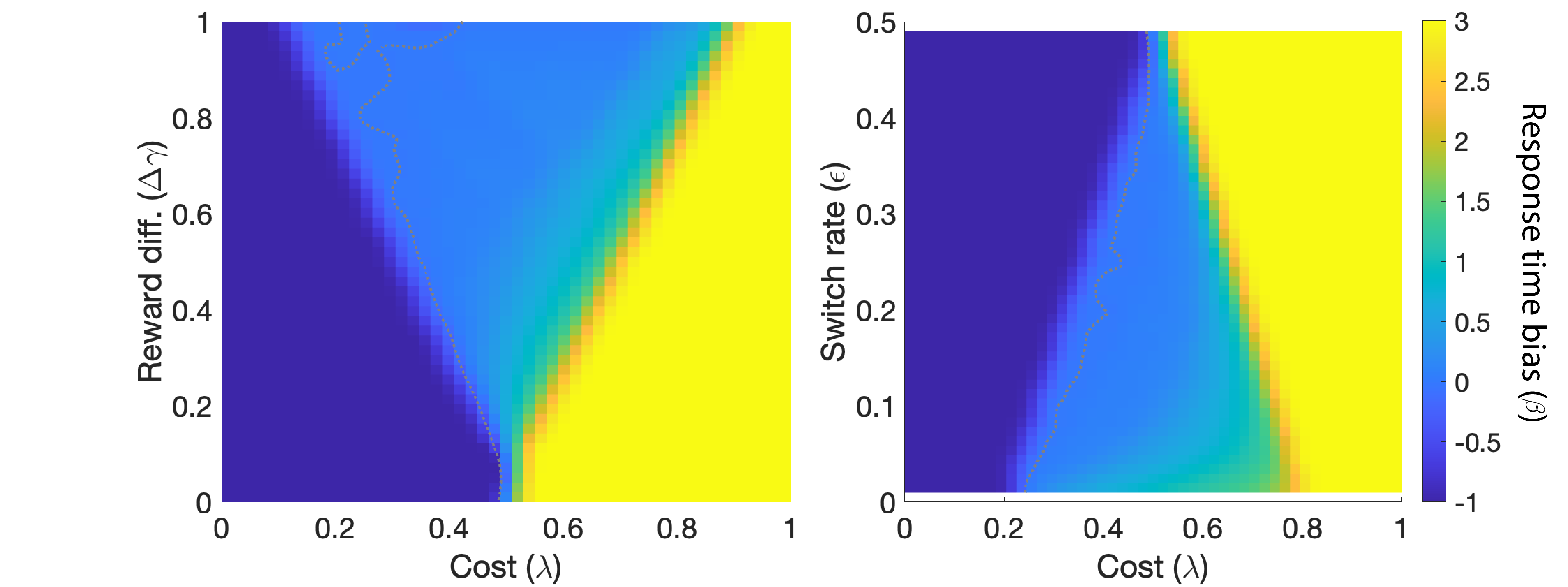}
    \caption{
    \textbf{Asymmetry in collective response times.} 
    Collective responses are slower for low-to-high than high-to-low environmental switches.
    Response times for low-to-high and high-to-low transitions are denoted by $\tau_{-,+}$ and $\tau_{+,-}$, respectively, and are estimated as mean response times from Monte Carlo simulations (with unsuccessful responses counted as inter-switch intervals).
    Shown is the response-time bias $\beta = \log(\tau_{-,+}/\tau_{+,-})$ across variation in environmental parameters:
    \textbf{(A)} Commitment cost $\lambda$ vs. reward difference $\Delta\gamma$;
    \textbf{(B)} Commitment cost $\lambda$ vs. environmental switching rate $\epsilon$.
    The gray-dashed curve shows the contour $\beta = 0$, corresponding to equal response times for low-to-high and high-to-low environmental switches.
    At low commitment cost, strong commitment (in low reward state) enables rapid responses to low-to-high switches, whereas at high cost weak commitment impedes timely responses to high-to-low switches.
    }
\end{figure}

\begin{figure}
    \centering
    \includegraphics[width=0.8\linewidth]{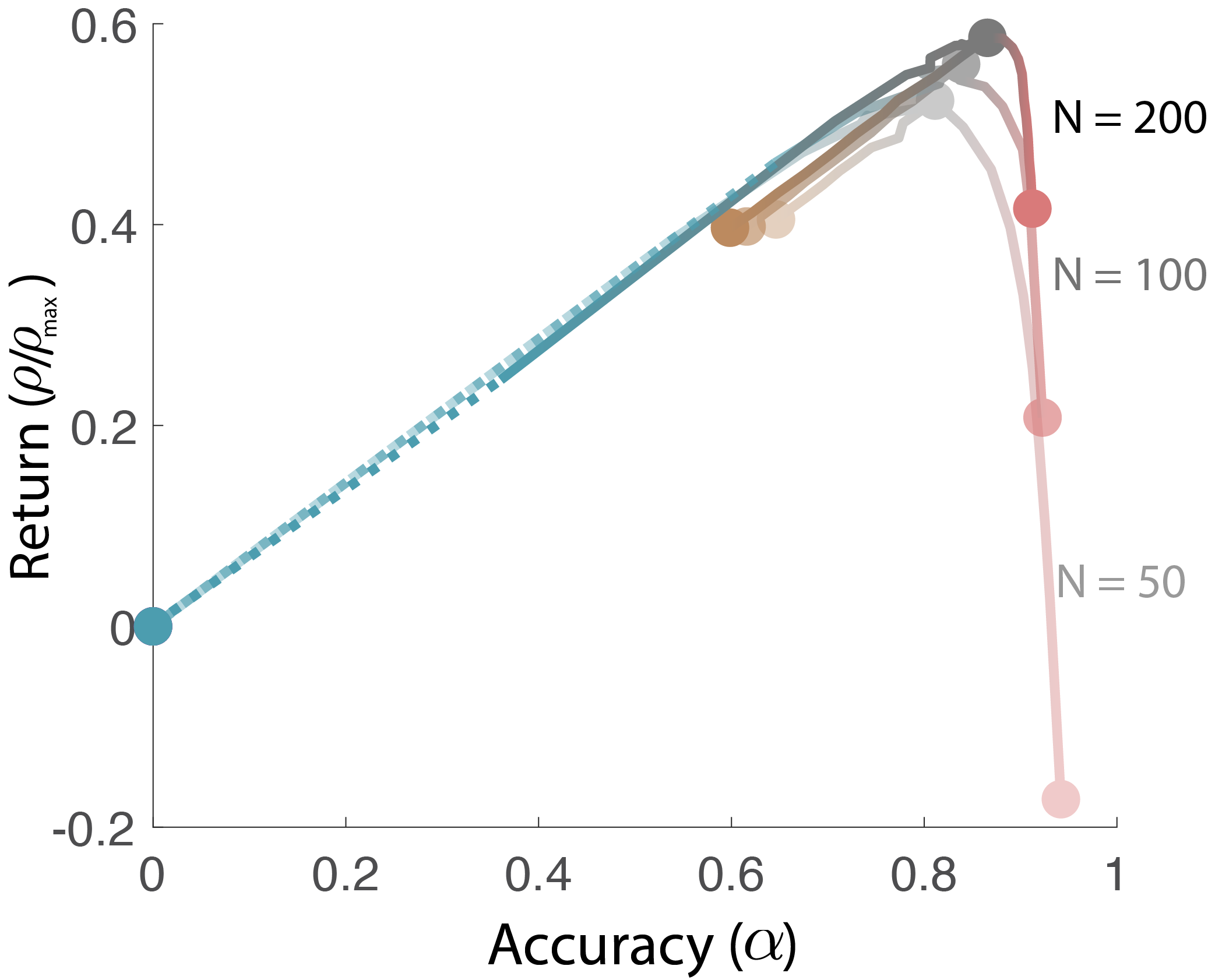}
    \caption{
    \textbf{Group size effect on collective performance under various perturbation on optimal threshold structure.} 
    The optimal performance increases in group size $N$:
    \textbf{(A)} $N = 50$,
    \textbf{(B)} $N = 100$,
    \textbf{(C)} $N = 200$.
    As group size increases, the normalized return increases under the same perturbations. Since the optimal fraction of risk-tolerant individuals decreases in larger groups, assigning maximal risk to the risk-tolerant individuals takes a lower relative cost than in small groups. 
    All other parameters are identical to those in Fig. 3.
    }
\end{figure}

\begin{figure}
    \centering
    \includegraphics[width=\linewidth]{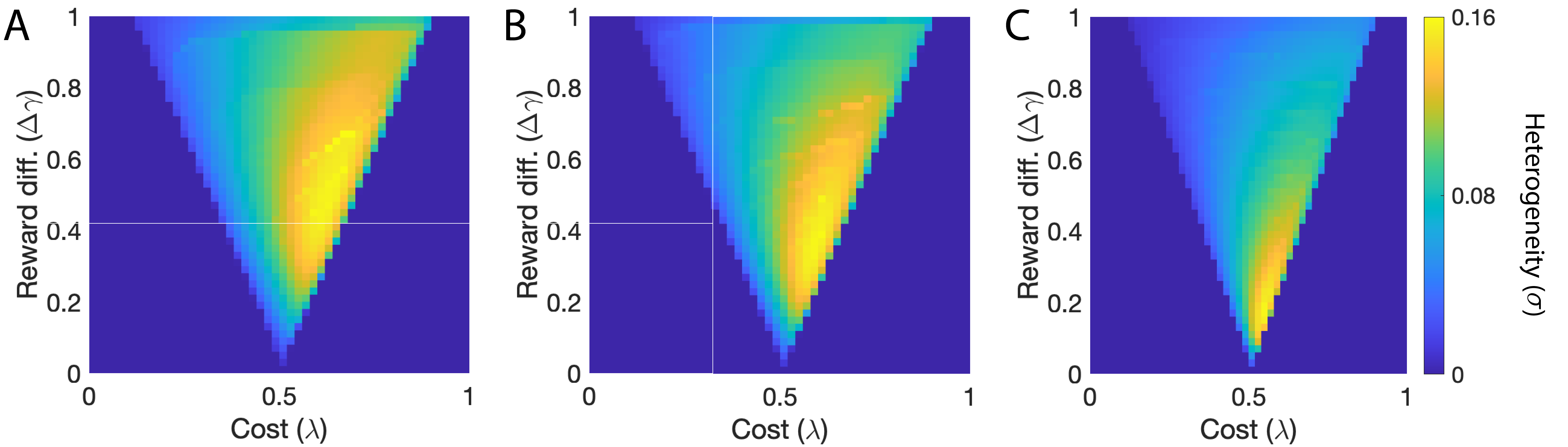}
    \caption{
    \textbf{Group size effect on optimal heterogeneity.} 
    Heterogeneity of the optimal threshold structure in decentralized collectives decreases with group size $N$: 
    \textbf{(A)} $N = 50$,
    \textbf{(B)} $N = 100$,
    \textbf{(C)} $N = 200$.
    All other parameters are identical to those in Fig. 4.
    }
\end{figure}

\end{document}